\documentclass[aps,prd,groupedaddress,showpacs]{revtex4}

\usepackage[parfill]{parskip}    
\usepackage{graphicx}
\usepackage{amssymb}
\usepackage{epstopdf}
\usepackage{color}
\usepackage{float}
\usepackage{amsmath}

\usepackage{graphicx}
\usepackage{pdfpages}
\usepackage[colorlinks=true,citecolor=blue,urlcolor=magenta,breaklinks]{hyperref}

\begin{document}

\title{ Charged anisotropic compact objects obeying Karmarkar condition }

\author{Y. Gomez-Leyton}
\email{ygomez@ucn.cl}
\affiliation{Departamento de F\'isica, Universidad Cat\'olica del Norte, Av. Angamos 0610, Antofagasta,
Chile.}

\author{Hina Javaid}
\email{Hinajavaid709@gmail.com}
\affiliation{Comsats University Islamabad, Lahore Campus, 51040, Pakistan.}

\author{L. S. Rocha}
\email{livia.silva.rocha@usp.br}
\affiliation{Universidade de S\~{a}o Pablo, Rua do Mat\~{a}o, 1226--Butant\~{a}, S\~{a}o Pablo--SP, 03178--200, Brazil. }

\author{Francisco Tello-Ortiz}
\email{francisco.tello@ua.cl}
\affiliation{Departamento de F\'isica, Facultad de ciencias básicas, Universidad de Antofagasta, Casilla 170, Antofagasta, Chile.}


\begin{abstract}
This research develops a well--established analytical solution of the Einstein-Maxwell field equations. We analyze the behavior of a spherically symmetric and static interior driven by a charged anisotropic matter distribution. The class I methodology is used to close the system of equations and a suitable relation between the anisotropy factor and the electric field is imposed. The inner geometry of this toy model is described using an ansatz for the radial metric potential corresponding to the well--known isotropic Buchdahl space-time. The main properties are explored in order to determine if the obtained model is appropriate to represent a real compact body such as neutron or quark star. {We have fixed the mass and radii using the data of the compact objects} SMC X--1 and LMC X--4. It was found that the electric field and electric charge have magnitudes of the order of $\sim 10^{21}\ [V/cm]$ and $\sim 10^{20}\ [C]$, respectively. The magnitude of the electric field and electric charge depends on the dimensionless parameter $\chi$. To observe these effects on the total mass, mass--radius ratio and surface gravitational red--shift, we computed numerical data for different values of $\chi$.   

\end{abstract}

\maketitle

\section{Introduction}\label{sec1}

It is well-known that any $n$-dimensional Riemaniann variety $V_ {n}$, can be embedded in a $m$-dimensional pseudo-Euclidean space $E_{m}$, whose dimension is given by $m=n\left(n+1\right)/2$ \cite{eisland,eisenhart}. The minimum extra dimension of the pseudo-Euclidean space is {$ \text{dim} (V_{n}) \leq \, m-n = n \, \left (n-1 \right)/ 2$}, which constitutes the class of the embedded space. On the other hand, some known important solutions in the arena of General Relativity (GR) are for example the Kerr space-time which corresponds to a class $p=5$ \cite{kerr}. Exterior and  interior Schwarzschild \cite{schwar} solutions are associated with classes $p=2$ and $p=1$, respectively, while the  Friedman–Lemaitre-Robertson–Walker space-time \cite{robert} is of class $ p= 1$. In this regard, the so-called class I condition \cite{karmarkar,sharma} has become a versatile tool to find out spherically symmetric solutions of the Einstein field equations describing the behavior of real compact structures as white dwarf, neutron and quark stars, driven by anisotropic (charged/uncharged) matter distributions \cite{r3,r5,r6, r10,r16,r17,r18,r19,r20,r21,r22,r23,r24,r25,r26,r27,r28,r32,r33,r34,r35,r36,r37,r38,tello1,r39} (and references contained therein). In a more widely context, this methodology was used to explain the existence of dark matter \cite{r40} based on the extra dimension argument. Moreover, the class $I$ scheme spreads into arena of modified gravity theory to study the existence of compact objects \cite{r41,r42}. Recently, this methodology was generalized to obtain all possible spherically symmetric structures satisfying Karmarkar condition \cite{r43} and extended into the gravitational decoupling by minimal geometric deformation \cite{r44}.

Despite the plethora of {articles addressing the construction of } compact configurations describing real astrophysical systems, determine {admissible models satisfying Einstein field equations} remains a great challenge. For the simplest cases the material content filling the stellar interior have an isotropic fluid distribution \i.e, $p_{r}=p_{\perp}$ \cite{lake}. Relaxing the stringent condition ($p_{r}=p_{\perp}$) to allow local anisotropies $\Delta\equiv p_{\perp}-p_{r}$ in the stellar medium constitutes a more realistic situation from the astrophysical point of view. 
{In this respect Ruderman \cite{n1}}, Canuto \cite{n2, n3,n4} and Canuto et. al \cite{n5,n6,n7,n8} investigations revealed that, if the density of matter {overcomes the} nuclear density, this problem is anisotropic in nature and should be treated in a relativistic way. In this direction, Bowers and Liang \cite{n9} presented one of the first studies based on  anisotropic matter distributions. Furthermore, some works predicted that anisotropic matter distributions can occur due to the existence of a solid stellar nucleus, by phase transitions and pion condensation \cite{n10}, the presence of type IIIA superfluid \cite{n11,n12}, the presence of rotation and electromagnetic field in the system \cite{n13,n14,n15}, among others.

The model proposed by Heintzmann and Hillebrandt \cite{a1} for neutron stars showed that for cases with a large and arbitrary anisotropy there is no mass limit for the stars. However, it has been determined that the maximum mass of a neutron star is still beyond  $3-4 M_{\odot}$ \cite{n16} so that the equilibrium is maintained. The theoretical possibility of anisotropy in strange stars was considered, with densities greater than those in neutron stars but lower than those in black holes.

The chances of having anisotropy in a compact star increase due to the relativistic interaction between the constituent particles and the random movements that they generate, resulting in a break in the uniformity of the distribution of the entire region. The functional foundations on which these types of structures are based are well understood and have been extensively studied in \cite{herre,harko,a2,a4,a5,a6,a7,a8,a9,a10,a11,a12,a13,a15,a16,a17,a19,a20,a21,a22}. To introduce anisotropy in the matter distribution allows adding important benefits in the description of the system, highlighting: $i)$ the presence of an extra gradient, repulsive in nature when $\Delta>0$ (otherwise is attractive). {This fact is relevant since the presence of a repulsive anisotropy gradient offset the gravitational attraction avoiding a gravitational collapse}, $ii)$ the possibility to obtain more compact objects and finally, $iii)$ the stability of the system is enhanced.

{Besides,} the matter distribution might contain additional ingredients to further improve the mentioned properties. For example, the inclusion of electrical features help to {balance} the gravitational gradient, improving the stability and balance of the system  \cite{RayY}. Moreover, the gravitational mass acquires an extra portion provided by the electrical contribution. Since the pioneering works by Bonnor \cite{e1} and Rosseland \cite{rose}, the effects of electrical properties on self-gravitating compact objects have been widely explored \cite{e2, e3, e4, e5, e6, e7, e8, e9, e10, e11, e12} (and references contained therein). As it is well known, Newton's theory of gravitation and electrostatics admit charged fluid  configurations in equilibrium, where the charge density can be as large as the mass density, in appropriate units. In a particular case the presence of electrons in stellar matter is crucial for the possible existence of a nuclear crust in quark--like stars, where electron displacement is associated with an electric current. {In this regard, a star like the Sun is expected to have a net electric charge due to the escape of electrons rather than the escape of photons. }

Following these good antecedents, in the present work we obtain an analytical charged anisotropic solution of the Einstein--Maxwell field equations for a spherically symmetric and static space-time. To build the inner manifold we have used the so-called class I condition. As this condition links both potentials, the $g_{rr}$ metric component is taken to be the Buchdahl metric potential \cite{buch}, then $g_{tt}$ is obtained. With the geometrical description at hand, it is imposed a relation between the anisotropy factor and the electric field \cite{dey} to close the system. The relation is controlled by a dimensionless parameter $\chi$. This constraint is well motivated since both quantities are null at the center of the configuration an increasing functions with increasing radial coordinate. An interesting point deduced from this link is that the anisotropy factor increases in magnitude with increasing $\chi$, whereas the electric field has the opposite behaviour. {The effects of the electric characteristics on the matter distribution, are evident when $\chi$ grows up in magnitude.} The structure of the problem naturally discards negative values for $\chi$, because for $\chi<0$ the radial pressure dominates the tangential one, then $\Delta<0$, which implies an unstable system.

To support our analysis we have used the mass and radii  of the real compact objects SMC X--1 and LMC X--4 \cite{r47}. Highlighting that quarks are small fundamental fragments of matter, which are interestingly combined to form the heaviest particles \cite{Ivanenko}, constituted by the three lightest flavors (top, bottom, and strange quarks \cite{Farhi}). These elementary particles refuse to be seen individually, and the physics associated with grouped quarks is quite complex. In the framework of stellar interiors, these types of structures have extremely high pressure and temperature, forcing the nuclear constituents to form quark matter or strange matter. Currently, there is no solid scientific basis to confirm or reject this hypothesis and thus explain various astrophysical phenomena. Nevertheless, from the phenomenological and theoretical point of view the study of quark or strange stars is fundamental to understand the fast radio burst process. Despite the possibility is still theoretical, there are good antecedents confirming that the collapse of these strange star crusts could be the origin of fast radio bursts mechanism \cite{l1,l2}.

To verify the viability of the proposed model, a complete graph is made, studying the thermodynamic variables, the electrical properties, the hydrostatic balance and the stability using a relativistic adiabatic index, subliminal sound velocities and the Harrison--Zeldovich--Novikov procedure. In addition, we discussed the impact of the electric field on macro-physical observables, such as total gravitational mass, compactness factor, and gravitational red surface displacement.

The document is organized as follows: The Sec. \ref{sec2}  is dedicated to presenting the Class I methodology, The Einstein--Maxwell field equations and the proposed model. In Sec. \ref{sec3} the toy model is matched in a smooth way with the exterior space-time described by the Reissner-Nordstr\"{o}m solution, in order to determine the space parameter that characterizes the model. Sec. \ref{sec4} is devoted to physical and mathematical analysis of the main features, such as the thermodynamic variables, energy conditions and electric properties. In Sec. \ref{sec5} the hydrostatic balance and stability analysis is performed to check the reliability of the solution. Sec. \ref{sec6} discuss the impact of electric component on the main macro-physical observables mentioned before. Finally, Sec. \ref{sec7} provides some remarks and conclusions for the reported study. It worth mentioning that throughout the article we shall employ relativistic geometrized units \i.e, $G=c=1$, the mostly negative signature $\{+,-,-,-\}$ and the following definition of the Riemann tensor: $R^{\gamma}_{\ \mu\nu\rho}=\partial_{\nu}\Gamma^{\gamma}_{\ \mu\rho}-\partial_{\mu}\Gamma^{\gamma}_{\ \nu\rho}+\Gamma^{\gamma}_{\ \mu\lambda}\Gamma^{\lambda}_{\ \nu\rho}-\Gamma^{\gamma}_{\ \nu\lambda}\Gamma^{\lambda}_{\ \mu\rho}$.

\section{Class I Approach Revisited and The Toy Model}\label{sec2}
In this section, we provide a detailed revision about class I condition and how it works in the framework of GR. Let us start by describing the usual mathematical complications in dealing with Einstein--Maxwell field equations to find out solutions describing compact objects such as neutron stars. Then, class I approach is presented as an auxiliary condition coming from purely geometric considerations, to reduce the mathematical issues and finally solve the system of equations. The salient toy model representing compact structures is also presented.  

\subsection{Class I Methodology}

{Since Einstein field equations constitute a complex set of couple partial non--linear differential equations, given by}  
\begin{equation}\label{eq1}
R_{\mu\,\nu}-\frac{R}{2}\,g_{\mu\,\nu}=8\,\pi\, T_{\mu\,\nu},
\end{equation}
finding solutions that satisfy this intricate system of coupled equations is very difficult, whether the problem is analyzed analytically or numerically. In the simplest case, when the geometry of the space-time is described by a spherically symmetric and static line element, expressed in canonical coordinates $x^{\mu}\equiv\left(t,r,\theta,\phi\right)$ as
\begin{equation}\label{eq2}
ds^{2}=e^{\eta}\,dt^{2}-e^{\lambda}\,dr^{2}-r^{2}\,d\Omega^{2},
\end{equation}
where $\left[\eta=\eta\left(r\right),\lambda=\lambda\left(r\right)\right]$ \i.e, purely radial functions, the problem is greatly reduced from the mathematical point of view. Nevertheless, the complexity in solve this set of equations also depends on the form of the matter distribution. {In this concern, we shall assume an imperfect fluid coupled to an electromagnetic field 
\begin{equation}\label{eq3}
T^{\mu}_{\nu}={M}^{\mu}_{\nu}+E^{\mu}_{\nu}=\underbrace{\left({\rho} +{\bar{p}}_{\perp}\right)\,\chi^{\mu}\,\chi_{\nu}-\delta^{\mu}_{\nu}\,{\bar{p}}_{\perp}+\left({\bar{p}}_{r}-{\bar{p}}_{\perp}\right)\,u^{\mu}\,u_{\nu}}_{M^{\mu}_{\nu}}+\underbrace{\left(-F^{\mu\,\alpha}F_{\nu\,\alpha}+\delta^{\mu}_{\nu}F_{\beta\,\alpha}\,F^{\beta\,\alpha}/4\right)/4\,\pi}_{E^{\mu}_{\nu}},
\end{equation}
with $\rho$ being the energy density, and ${\bar{p}}_{r}$ and ${\bar{p}}_{\perp}$ being the pressure waves {along the main directions of the fluid sphere \i.e, the radial and transverse ones, respectively. The time--like vector $\chi^{\omega}=e^{-\eta/2}\delta^{\omega}_{\ t}$, satisfying $\chi^{\omega}\chi_{\omega}=1$, represents the velocity of the fluid}. Moreover $u^{\omega}=e^{-\lambda/2}\delta^{\omega}_{\ r}$ is a unit space-like vector in the radial direction (orthogonal to $\chi^{\omega}$), satisfying $u^{\omega}u_{\omega}=-1$. The tensor $F_{\mu\nu}$, represents the skew--symmetric electromagnetic tensor defined as usual by
\begin{equation}\label{a01}
F_{\mu\nu}=\partial_{\mu}A_{\nu}-\partial_{\nu}A_{\mu}   
\end{equation}
and in the present context
satisfying the covariant vacuum Maxwell equations
\begin{equation}\label{eq4}
\nabla_{[\alpha}F_{\beta\gamma]}=0, \quad \mbox{and} \quad \nabla_{\beta}\,F^{\beta\, \alpha}=4\,\pi \, J^{\alpha},
\end{equation}
being $J^{\alpha}$ the four electric current. In the static and spherically symmetric case,
the vectorial potential $A^{\mu}$ is expressed as follows
\begin{equation}\label{a1}
A^{\mu}=\left(\Phi(r),0,0,0\right)=\Phi(r)\delta^{\mu}_{\ t},    
\end{equation}
where $\Phi(r)$ is the electric scalar potential, and the four electric current is expressed by
\begin{equation}\label{a2}
J^{\mu}=\left(e^{-\nu/2}\,\sigma,0,0,0\right)=e^{-\nu/2}\,\sigma\,\delta^{\mu}_{\ t},    
\end{equation}
where $\sigma$ is representing the surface charge density.
Now integrating the expression at the left hand side in (\ref{eq4}) one gets
\begin{equation}\label{a3}
F^{tr}(r)=-F^{rt}(r)=\frac{q(r)}{r^{2}}\,e^{-(\lambda(r)+\nu(r))/2}.  
\end{equation}
As usual the electric charge $q(r)$ has been defined by using the relativistic Gauss's law as follows \cite{bek,dio}
\begin{equation}\label{eq5}
 q\left(r\right)=4\pi\int^{r}_{0}\sigma(x)x^{2}e^{\lambda(x)/2}dx=r^{2}\,\sqrt{-F^{tr}F_{tr}}.
\end{equation}
Furthermore, as we are dealing with a spherically symmetric and static configuration, this implies that the only non--vanishing components of the electromagnetic tensor $F^{\mu\nu}$ are $F^{tr}=-F^{rt}$. These components are just the electric field $E=E(r)$ along the radial direction. Thus, from Eq. (\ref{a3}) one obtains
\begin{equation}\label{a6}
E(r)=\frac{q(r)}{r^{2}}\, e^{-(\lambda(r)+\nu(r))/2}.    
\end{equation}
Thus the energy--momentum tensor (\ref{eq3}) can be cast as follows
\begin{equation}\label{t}
T^{\mu}_{\nu}=\text{diag}\left(\rho+\frac{E^{2}}{8\pi},-p_{r}+\frac{E^{2}}{8\pi},-p_{\perp}-\frac{E^{2}}{8\pi},-p_{\perp}-\frac{E^{2}}{8\pi}\right).    
\end{equation}
It should be noted that the form of the energy--momentum tensor (\ref{eq3}) arises from the variation of the full action (gravitational field minimally coupled to the electromagnetic field, source and particles) with respect to the metric tensor. For further details about this point see appendix \ref{A}.}

{Next, putting together Eqs. (\ref{eq1})--(\ref{eq3}), one obtains the following set of equations \cite{bek,dio}
\begin{eqnarray}\label{eq6}
8\,\pi \,{\rho}+E^{2}&=&\frac{1}{r^2}-e^{-\lambda}\left(\frac{1}{r^2}-\frac{\lambda^{\prime}}{r}\right),\\\label{eq7}
8\,\pi \,{p}_{r}-E^{2}&=&-\frac{1}{r^2}+e^{-\lambda}\left(\frac{1}{r^2}+\frac{\eta^{\prime}}{r}\right),\\\label{eq8}
8\,\pi \, {p}_{\perp}+E^{2}&=&\frac{1}{4}e^{-\lambda}\left(2\eta^{\prime\prime}+\eta^{\prime2}-\lambda^{\prime}\eta^{\prime}+2\frac{\eta^{\prime}-\lambda^{\prime}}{r}\right).
\end{eqnarray}}

It is evident that the above system (\ref{eq6})--(\ref{eq8}) contains six unknowns namely, the geometry $\{\eta,\lambda\}$, the thermodynamic variables $\{\rho,p_{r},p_{t}\}$ and the electric field $E$, but only three equations. The conservation of the energy--momentum tensor, 
\begin{equation}\label{eq9}
\nabla_{\mu}\,T^{\mu}_{\ \nu}=0 \,\Rightarrow -p^{\prime}_{r}-\frac{\eta^{\prime}}{2}\left(\rho+p_{r}\right)+\frac{2}{r}\left(p_{\perp}-p_{r}\right)+\sigma\, E \, e^{\lambda/2}=0,
\end{equation}
does not provide any additional information, since (\ref{eq9}) is a linear combination of Eqs. (\ref{eq6})--(\ref{eq8}). Is in this sense that the so-called class I condition acts as an auxiliary constraint to solve the given set of equations, although in this case to close the system it is necessary to impose for example an equation of state or another relation to determine the full energy--momentum tensor. This is due to the electric contribution. Nevertheless, if $E=0$ one recovers the usual Einstein equations for an anisotropic fluid distribution. Therefore, the class I condition is enough to close the problem at least from the mathematical point of view. \\ 

In general the 4--dimensional manifold given by Eq. (\ref{eq2}) corresponds to a space--time of class II. This means that is necessary a 6--dimensional pseudo--Euclidean space to encrust it. However, under a suitable parametrization the space--time (\ref{eq2}) can be immersed into a 5--dimensional pseudo-Euclidean space, turning into a class I \cite{eisland,eisenhart,karmarkar}. In general any variety $V_{n}$ can be embedded in a flat space of $n\left(n+1\right)/2$ dimensions \cite{eisland,eisenhart}. Nevertheless, if the lowest order of this flat space is $n+p$, we say that $V_{n}$ is of class $p$. Necessary and sufficient conditions must be satisfied for any spherically symmetric space--time (static and non--static) so it can be assigned as class I. Those are:
\begin{itemize}
    \item A system of symmetric quantities $b_{\mu\,\nu}$ must be established, such that
    \begin{equation}\label{eq10}
    R_{\mu\,\nu\,\alpha\,\beta}=\epsilon\,\left(b_{\mu\,\alpha}\,b_{\nu\,\beta}-b_{\mu\,\beta}\,b_{\nu\,\alpha}\right)\quad \left(\text{Gauss's equation}\right),
    \end{equation}
    where $\epsilon=\pm1$ whenever the normal to the manifold is space--like (+1)
or time--like (-1).
\item The system $b_{\mu\nu}$ must satisfy the differential equations
\begin{equation}\label{eq11}
\nabla_{\alpha}b_{\mu\,\nu}-\nabla_{\nu}\,b_{\mu\,\alpha}=0 \quad \left(\text{Codazzi's equation}\right).
\end{equation}
In general Codazzi's equation (\ref{eq11}) is not zero. This form of Eq. (\ref{eq11}) is implied by Eq. (\ref{eq10}) \cite{eisenhart}.
\end{itemize}

 From (\ref{eq2}) {the non vanishing elements} of the Riemann tensor are:
 \begin{eqnarray}\nonumber
 R_{trtr}&=&-\frac{e^{\eta}}{4}\left[2\,\eta^{\prime\prime}+\eta^{\prime \ 2}-\lambda^{\prime}\,\eta^{\prime}\right], \\ \nonumber
 R_{\theta\,\phi\,\theta\,\phi}&=& -e^{-\lambda}\,r^{2}\,sin^{2}\theta\left(e^{-\lambda}-1\right), \\ \label{eq12}
 R_{r\,\phi\, r\,\phi}&=&sin^{2}\theta\, R_{r\,\theta \,r\,\theta}=\frac{r}{2}\,\lambda^{\prime}, \\ \nonumber
 R_{t\,\theta\, t\,\theta}&=&sin^{2}\theta \,R_{t\,\phi\, t\,\phi}=-\frac{r}{2}\,\eta^{\prime}\,e^{\eta-\lambda}.
 \end{eqnarray}
So, by using the set of equations (\ref{eq12}) into Eq. (\ref{eq10}) one gets:
\begin{eqnarray}&&\label{eq13}
b_{tr}\,b_{\phi\,\phi}=R_{r\phi t\phi}=0;~~
b_{tr}\,b_{\theta\,\theta}=R_{r\theta t\theta}=0;\\ \nonumber&&
b_{tt}\,b_{\phi\,\phi}=R_{t\phi \,t\,\phi};~~
b_{tt}\,b_{\theta\,\theta}=R_{t\theta t\theta}; ~~
b_{rr}\,b_{\phi\,\phi}=R_{r\phi r\phi}; \\ \nonumber&&
b_{\theta\,\theta}\,b_{\phi\,\phi}=R_{\theta\,\phi\,\theta\,\phi}; ~~
b_{rr}\,b_{\theta\,\theta}=R_{r\theta\, r\theta}; ~~
b_{tt}\,b_{rr}=R_{trtr}.
\end{eqnarray} 

The above relations leads to:
\begin{equation}\label{eq14}
\left(b_{tt}\right)^{2}=\frac{\left(R_{t\,\theta\, t\,\theta}\right)^{2}}{R_{\theta\,\phi\,\theta\,\phi}}\,sin^{2}\theta, \quad \left(b_{rr}\right)^{2}=\frac{\left(R_{r\,\theta \,r\,\theta}\right)^{2}}{R_{\theta\,\phi\,\theta\,\phi}}\,sin^{2}\theta, \quad \left(b_{\theta\,\theta}\right)^{2}=\frac{R_{\theta\,\phi\,\theta\,\phi}}{sin^{2}\theta}, \quad \left(b_{\phi\,\phi}\right)^{2}=sin^{2}\theta \,R_{\theta\,\phi\,\theta\,\phi}.
\end{equation}
Upon replacing (\ref{eq14}) into  expression (\ref{eq13}) one gets:
\begin{equation}\label{eq15}
R_{t\,\theta\, t\,\theta}\,R_{r\,\phi\, r\,\phi}=R_{trtr}\,R_{\theta\,\phi\,\theta\,\phi},    
\end{equation}
subject to $R_{\theta\,\phi\,\theta\,\phi}\neq\,0$ \cite{sharma}. It should be noted that eq. (\ref{eq13}) satisfies Codazzi's equation (\ref{eq11}). On the other hand, in the case of a general non--static spherically symmetric space--time, the second and last equality in (\ref{eq13}) become:
\begin{equation}\label{eq16}
b_{tr}\,b_{\theta\,\theta}=R_{r\,\theta \,t\,\theta} \quad \mbox{and}\quad  b_{tt}\,b_{rr}-\left(b_{tr}\right)^{2}=R_{trtr},  
\end{equation}
where $\left(b_{tr}\right)^{2}=sin^{2}\theta \left(R_{r\theta t \theta}\right)^{2}/R_{\theta\,\phi\,\theta\,\phi}$. So, the class I condition becomes \cite{eisland,karmarkar}:
\begin{equation}\label{eq17}
R_{t\,\theta\, t\,\theta}\,R_{r\,\phi\, r\,\phi}=R_{trtr}\,R_{\theta\,\phi\,\theta\,\phi}+R_{r\theta t \theta}\,R_{r\,\phi\, t\,\phi}.    
\end{equation}
In this particular case, where the space--time is given by eq. (\ref{eq2}) the condition (\ref{eq15}) (or equivalently (\ref{eq17})) leads to:
\begin{equation}\label{eq18}
2\frac{\eta^{\prime\prime}}{\eta^{\prime}}+\eta^{\prime}=\frac{\lambda^{\prime}\,e^{\lambda}}{e^{\lambda}-1},   
\end{equation}
with $e^{\lambda}\neq 1$. This equation can be solved to express $\eta=\eta\left(\lambda\right)$ or $\lambda=\lambda\left(\eta\right)$. The results are:
\begin{equation}\label{eq19}
e^{\lambda}=1+C\,\eta^{\prime \, 2}\,e^{\eta},    
\end{equation}
or
\begin{equation}\label{eq20}
e^{\eta}=\left[A+B\int\,\sqrt{\left(e^{\lambda}-1\right)}\,dr\right]^{2},    
\end{equation}
being $\{A,B,C\}$ integration constants.
At this stage some comments are pertinent. As we pointed out before, the class I condition helps to reduce the mathematical complexity in order to tackle the set of equations (\ref{eq8})--(\ref{eq10}). This fact is reflected by Eq. (\ref{eq19}) or (\ref{eq20}), so with a suitable choice of one of the metric potentials the geometry of the space--time is completely determined. Therefore, with this information in hand the energy--momentum tensor is obtained from Eqs. (\ref{eq8})--(\ref{eq10}). It should be noted that the class I condition also serves to find out spherical solutions whose matter distribution is described by an isotropic fluid $p_{r}=p_{\perp}=p$. From Eqs. (\ref{eq9})--(\ref{eq10}) and (\ref{eq20}) the anisotropy factor $\Delta$ reads \cite{r3,r5,r6}:
\begin{equation}\label{eq21}
8\,\pi\,\Delta=\frac{\eta^{\prime}\,e^{-\lambda}}{4}\left(\frac{\eta^{\prime}e^{\eta}}{2\,B^{2}\,r}-1\right)\left(\frac{2}{r}-\frac{\lambda^{\prime}\,e^{-\lambda}}{1-e^{-\lambda}}\right)-2\,E^{2}.    
\end{equation}

Imposing the isotropic condition $\Delta=0$ and $E=0$ one obtains the following:
\begin{equation}\label{eq22}
\frac{\eta^{\prime}\,e^{-\lambda}}{4}\left(\frac{\eta^{\prime}\,e^{\eta}}{2\,B^{2}\,r}-1\right)\left(\frac{2}{r}-\frac{\lambda^{\prime}\,e^{-\lambda}}{1-e^{-\lambda}}\right)=0.    
\end{equation}
{It is observed from Eq. (\ref{eq22}), that} if the first parenthesis is zero, the corresponding solution will be the Kohlar--Chao \cite{koler} solution, while if the
second parenthesis is null, the interior Schwarzschild \cite{schwar} solution is obtained. Notwithstanding, if $E\neq0$, it is possible to build charged isotropic fluid spheres.  

\subsection{The Model}
As explained before, additional constraints are necessary to close the Einstein--Maxwell fields equations. To determine the geometrical sector we have imposed the $g_{rr}$ metric potential to be
\begin{equation}\label{eq23}
e^{\lambda(r)}=\frac{2\left(1+Cr^{2}\right)}{2-Cr^{2}}.    
\end{equation}
This metric potential corresponds to one of the solutions provided by Buchdahl \cite{buch}, and was already employed in the context of embedding class I methodology to build compact stars \cite{r10}. The motivation of this choice relies on the well physical and mathematical properties that it has \i.e, finite at $r=0$, positive defined and increasing function with increasing radial coordinate. These features, in principle\footnote{In this case the metric potential (\ref{eq23}) does not determine completely the behavior of the density $\rho$, because the electric field contribution plays and important role on the behaviour of $\rho$.} assure a well posed energy density $\rho$ within the stellar medium. So, by inserting Eq. (\ref{eq23}) into (\ref{eq20}) one arrives to
{
\begin{equation}\label{eq24}
e^{\eta(r)}=\left(A-\bar{B}\sqrt{2-Cr^{2}}\right)^{2},  \end{equation}
where $A$ and $\bar{B}$ are dimensionless parameters and $C$ has units of $\text{length}^{-2}$. In order to present the results in a more compact fashion, in Eq. (\ref{eq24}) we have defined $\bar{B}$ to be equal to $\sqrt{3}B/\sqrt{C}$.} Thus, the space--time describing the stellar interior is given by
\begin{equation}\label{eq25}
ds^{2}=\left(A-\bar{B}\sqrt{2-Cr^{2}}\right)^{2}dt^{2}-\frac{2\left(1+Cr^{2}\right)}{2-Cr^{2}}dr^{2}-r^{2}d\Omega^{2}.    
\end{equation}
To close the system (\ref{eq8})--(\ref{eq10}) we shall employ the following ansatz
\begin{equation}\label{eq26}
\Delta=2\chi E^{2},    
\end{equation}
where $\chi$ is a dimensionless constant. This relation was  recently employed to build higher dimensional fluid spheres in the background of Finch--Skea space--time \cite{dey}. The main physical reason behind (\ref{eq26}) is that both $\Delta$ and $E$ have the same behavior inside the compact structure \i.e, positive defined and increasing functions with increasing radial coordinate (as we will see later, these requirements are necessary to describe a well posed charged compact configuration), implying that at $r=0$: $\Delta(0)=E(0)=0$. {Now subtracting Eqs. (\ref{eq7}) from (\ref{eq8}) we obtain
\begin{equation}\label{new1}
8\pi\left(p_{\perp}-p_{r}\right)(r)+2E^{2}(r)=f(r).    
\end{equation}
Remembering that the anisotropy factor is defined as $\Delta\equiv p_{\perp}-p_{r}$, the Eq. (\ref{new1}) becomes
\begin{equation}\label{new2}
8\pi\Delta(r)+2E^{2}(r)=f(r).    
\end{equation}
Therefore, putting together (\ref{eq26}) and (\ref{new2}) one gets
\begin{equation}\label{eq27}
E^{2}(r)=\frac{f(r)}{2\left(1+8\pi\chi\right)}. 
\end{equation}}
The anisotropy factor is then expressed by
\begin{equation}\label{eq28}
\Delta(r)= \frac{f(r)\chi}{\left(1+8\pi\chi\right)}, 
\end{equation}
where 
\begin{equation}\label{eq29}
f(r)= \frac{e^{-\lambda}}{4r^{2}}\left[ 4 \left(e^{\lambda }-1\right)+r \left(\eta ' \left(r \eta '-2\right)-\lambda ' \left(2+r \eta
'\right)+2 r \eta ''\right)\right] =\frac{C^2 r^2 \left[4 \bar{B} \left(C r^2-2\right)+3 A \sqrt{2-C r^2}\right]}{2
\left[1+C r^2\right]^2 \left[\bar{B} \left(C r^2-2\right)+A \sqrt{2-C r^2}\right]}
\end{equation}
for the present model described by the metric potentials (\ref{eq23}) and (\ref{eq24}).
At this stage some comments are pertinent
\begin{enumerate}
    \item Equation (\ref{eq28}) restricts $\chi$ to be positive defined, since $\chi<0$ implies $\Delta(r)<0$, then the system will be unstable. 
    \item In the case $\chi=0$ one obtains an isotropic charged solution ($\Delta(r)=0$).
    \item If $E^{2}(r)=0$ one gets an anisotropic fluid ball with anisotropy factor given $\Delta(r)=f(r)/8\pi$.
\end{enumerate}
In what follows we shall assume $\chi\neq0$, $E^{2}(r)\neq0$ and $\Delta(r)\neq0$ with the aim to build a toy model representing a compact charged anisotropic fluid sphere. Replacing equations (\ref{eq23})--(\ref{eq24}) and (\ref{eq27}) into the system (\ref{eq6})--(\ref{eq8}) one yields at the following thermodynamic variables
\begin{eqnarray}\label{eq30}
\rho(r)&=&\frac{3 C \left[3+C r^2\right]}{16\pi \left[1+C r^2\right]^2}-\frac{E^{2}(r)}{8\pi}, \\ \label{eq31}
p_{r}(r)&=&\frac{C \left[5 \bar{B} \left(2-C r^2\right)-3 A \sqrt{2-C r^2}\right]}{16\pi \left[1+C
r^2\right] \left[\bar{B} \left(C r^2-2\right)+A \sqrt{2-C r^2}\right]}+\frac{E^{2}(r)}{8\pi}, \\ \label{eq32}
p_{\perp}(r)&=&\frac{C \left[\bar{B} \left(2-C r^2\right) \left(5+C r^2\right)-3 A \sqrt{2-C r^2}\right]}{16\pi
\left[1+C r^2\right]^2 \left[\bar{B} \left(C r^2-2\right)+A \sqrt{2-C r^2}\right]}-\frac{E^{2}(r)}{8\pi}.
\end{eqnarray}
In the next section the junction condition formalism is performed in order to determine the space parameter $\{A,\bar{B},C\}$ characterizing the model.

\section{Junction Conditions at the Star Surface }\label{sec3}

{As we are dealing with a finite configuration. In order to guaranteed that the toy model represented by (\ref{eq25}) which is describing the inner space--time $\mathcal{M}^{-}$ is well established, one needs to join it in a smoothly way at the surface interface $\Sigma: r=R$ (or equivalently the compact object surface), with external space--time $\mathcal{M}^{+}$.  In this case the exterior manifold is no longer empty since we are facing a charged compact configuration, thus exterior variety $\mathcal{M^{+}}$ is given by the Reissner--Nordstr\"{o}m solution}
\begin{equation}\label{eq33}
ds^{2}=\left(1-\frac{2\bar{M}}{r}+\frac{Q^{2}}{r^{2}}\right)dt^{2}-\left(1-\frac{2\bar{M}}{r}+\frac{Q^{2}}{r^{2}}\right)^{-1}dr^{2}-r^{2}d\Omega^{2}.    
\end{equation}
{This process known as the matching condition procedure, serves to obtain the complete set of constant parameters $\{A,\bar{B},C\}$. To do this, it is employed the Israel--Darmois (ID) \cite{is,dar} mechanism. This method is supported by the continuity of the temporal $g_{tt}$ and radial $g_{rr}$ metric potentials throughout
the surface interface $\Sigma$. Technically, this condition is known as the first fundamental form. Explicitly it reads}
\begin{equation}\label{eq34}
g^{-}_{tt}\bigg{|}_{r=R}=g^{+}_{tt}\bigg{|}_{r=R} \quad \mbox{and} \quad
g^{-}_{rr}\bigg{|}_{r=R}=g^{+}_{rr}\bigg{|}_{r=R}.
\end{equation}

In our scenario we have
\begin{eqnarray}\label{eq35}
\left(A-\bar{B}\sqrt{2-CR^{2}}\right)^{2}&=&1-\frac{2\bar{M}}{R}+\frac{Q^{2}}{R^{2}}, \\ \label{eq36}
 \frac{2-CR^{2}}{2\left(1+CR^{2}\right)}&=&1-\frac{2\bar{M}}{R}+\frac{Q^{2}}{R^{2}}.
\end{eqnarray}

{At junction surface $\Sigma$ both $M$ and $\bar{M}$ coincide. Thus, the total mass $M$ inside the charged ball is determined by the external space--time. As we are gluing the interior geometry $\mathcal{M}^{-}$ with outer one $\mathcal{M}^{+}$, at the surface $r=R$ is induced by $g^{-}_{\mu\nu}$ and $g^{+}_{\mu\nu}$ an intrinsic geometry described by the extrinsic curvature tensor $K_{ij}$. Precisely, the continuity of $K_{rr}$ across $\Sigma$, the so--called second fundamental form, guaranteed a completely null radial pressure}
\begin{equation}\label{eq37}\resizebox{0.94\hsize}{!}{
$
p_{r}(R)=0 \Rightarrow \frac{2C \left[3 A \sqrt{2-C R^2}+5 \bar{B} \left(C R^2-2\right)\right]\left[1+8\pi\chi\right]+C^2 R^2 \left[2 \bar{B} \left(C R^2-2\right)\left(3+40\pi\chi\right)+3 A \sqrt{2-C R^2}\left(1+16\pi\chi\right)\right]}{32\pi \left[1+C
R^2\right]^{2} \left[\bar{B} \left(C R^2-2\right)+A \sqrt{2-C R^2}\right]\left[1+8\pi\chi\right]}=0$}.    
\end{equation}

A null radial pressure at the boundary $\Sigma$ {is a necessary mechanism to confine the matter content inside a bound space--time region $0\leq r\leq R$, which in turn determines the size of the collapsed configuration \i.e, its radii $R$.  
Furthermore, the continuity  $K_{\phi\phi}$ and $K_{\varphi\varphi}$ fixes the total mass of the compact structure}
\begin{equation}\label{eq38}
\left[K^{-}_{\phi\phi}-K^{+}_{\phi\phi}\right]\bigg{|}_{\Sigma}=\left[K^{-}_{\varphi\varphi}-K^{+}_{\varphi\varphi}\right]\bigg{|}_{\Sigma}=0 \Rightarrow m(R)=M. 
\end{equation}
Besides, it is also necessary to impose the electric charge $q(r)$ continuity across the boundary $\Sigma$
\begin{equation}\label{eq39}
q(R)=Q=R^{2}E(R)=R^{2}\sqrt{\frac{C^2 R^2 \left[4 \bar{B} \left(C R^2-2\right)+3 A \sqrt{2-C R^2}\right]}{4\left[1+8\pi\chi\right]
\left[1+C R^2\right]^2 \left[\bar{B} \left(C R^2-2\right)+A \sqrt{2-C R^2}\right]}}.
\end{equation}

Equations (\ref{eq35})--(\ref{eq37}) and (\ref{eq39}) are the necessary conditions to close the problem and obtain the arbitrary constants $\{A,\bar{B},C\}$. In table \ref{table1} are exhibited the numerical values by the mentioned space parameter for total gravitational mass $M$ and radius $R$ corresponding with the compact stars SMC X--1 and LMC X--4 \cite{r47}. 

\begin{table}[H]
\caption{Numerical values of constant parameters $A$, $\bar{B}$ and $C$ for different values of $M$ and $R$ and $\chi=0.008$.  }
\label{table1}
\begin{tabular*}{\textwidth}{@{\extracolsep{\fill}}lrrrrrrrl@{}}
\hline
\multicolumn{1}{c}         {Strange Star} & 
\multicolumn{1}{c}         {$M/M_{\odot}$} & 
\multicolumn{1}{c}         {$R \ [km]$} &
\multicolumn{1}{c} {$C \ [\text{km}^{-2}]$}& 
\multicolumn{1}{c} {$A$\ (\text{Dimensionless)}} &
\multicolumn{1}{c} {$\bar{B}$\ (\text{Dimensionless})}\\
\hline
SMC X--1 ~\ \cite{r47} & 1.04&8.301&0.004436 &$-1.951878$&$-0.880693$\\
\hline
LMC X--4 ~\ \cite{r47} &1.29&8.831&$0.004761$&$-1.857516$& $-0.851659$    \\
\hline
\end{tabular*}
\end{table}

\section{On the physical and mathematical feasibility}\label{sec4}
{Here the principal physical and mathematical properties  representing the model are thoroughly analyzed. This point is carried out by using the usual analysis provided by Herrera et. al and Mak et. al \cite{herre,harko}.} 

\subsection{Geometry and Thermodynamic Description}

Let us start by analyzing the behavior of the inner geometry given by metric potentials (\ref{eq23})--(\ref{eq24}).
From the space--time generated by these potentials {we can remark} the following features:
\begin{itemize}
    \item {It is observed that, $e^{\eta}$ and $e^{\lambda}$ are free from pathologies \i.e, both are continuous functions everywhere, for all $r\in[0,R]$. }
    \item {At the center of the star, equation (\ref{eq24}) yields to $e^{\eta(0)}=B^{2}$ and equation (\ref{eq23}) to 
     $e^{\lambda(0)}=1$. This information corroborates that the geometry of the inner space--time is being described by two monotonically increasing metric functions whose minimum values are reached at $r=0$.  }      \end{itemize}
     
\begin{figure}[H]
\centering
\includegraphics[width=0.34\textwidth]{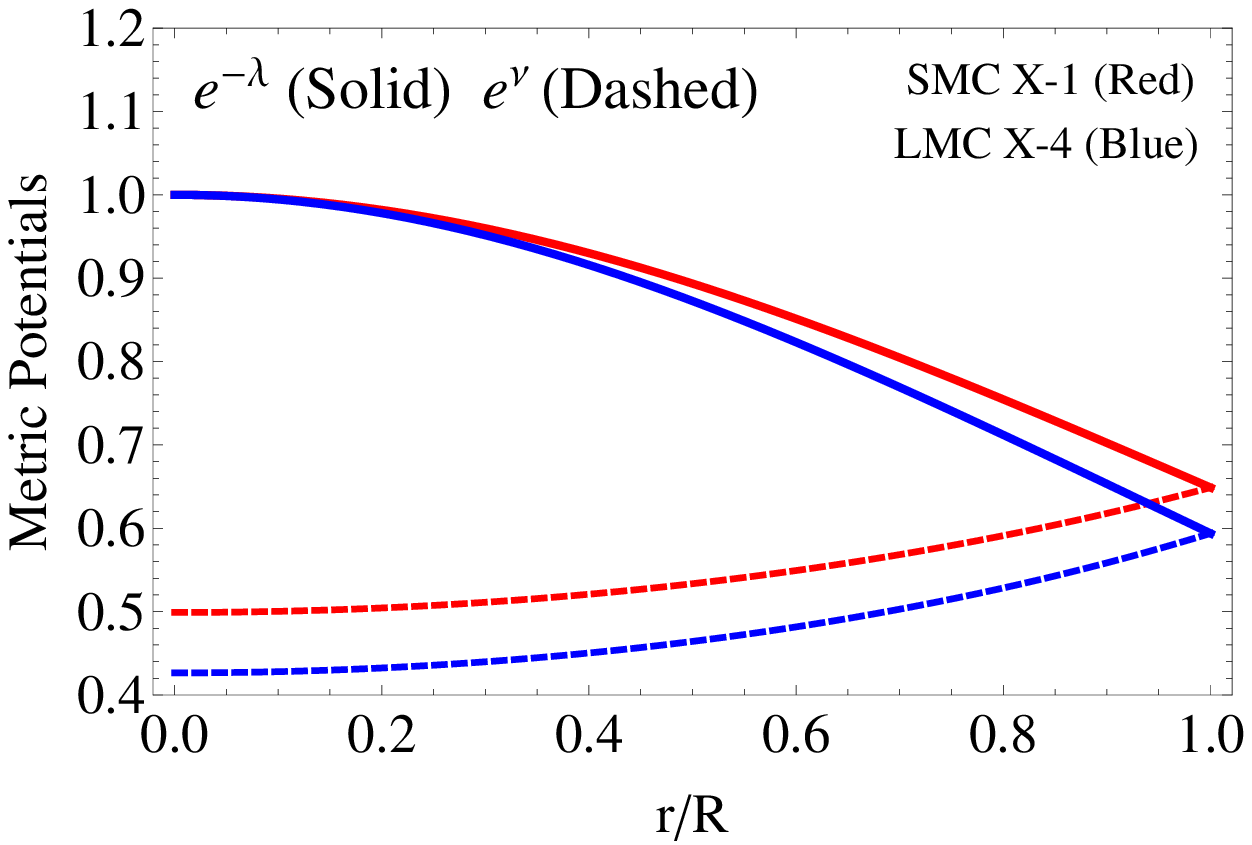}
\includegraphics[width=0.34\textwidth]{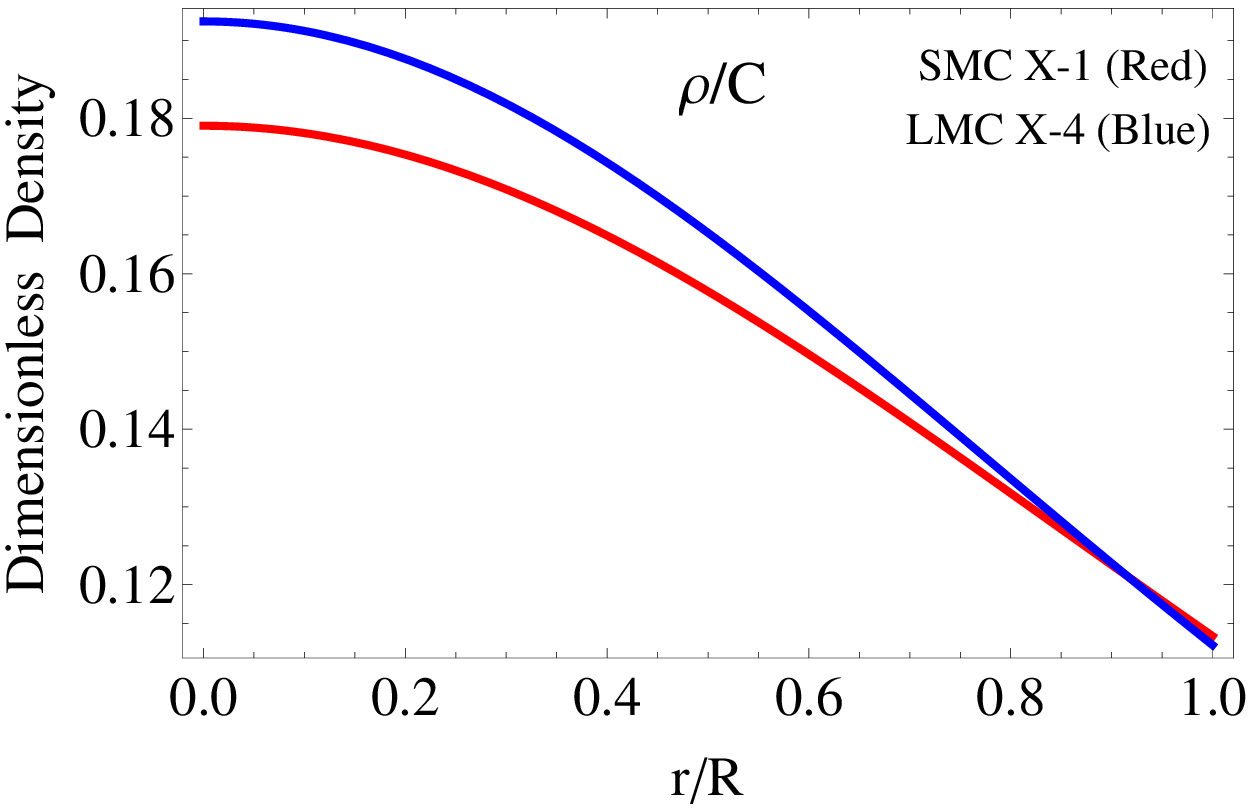}     \\
\includegraphics[width=0.34\textwidth]{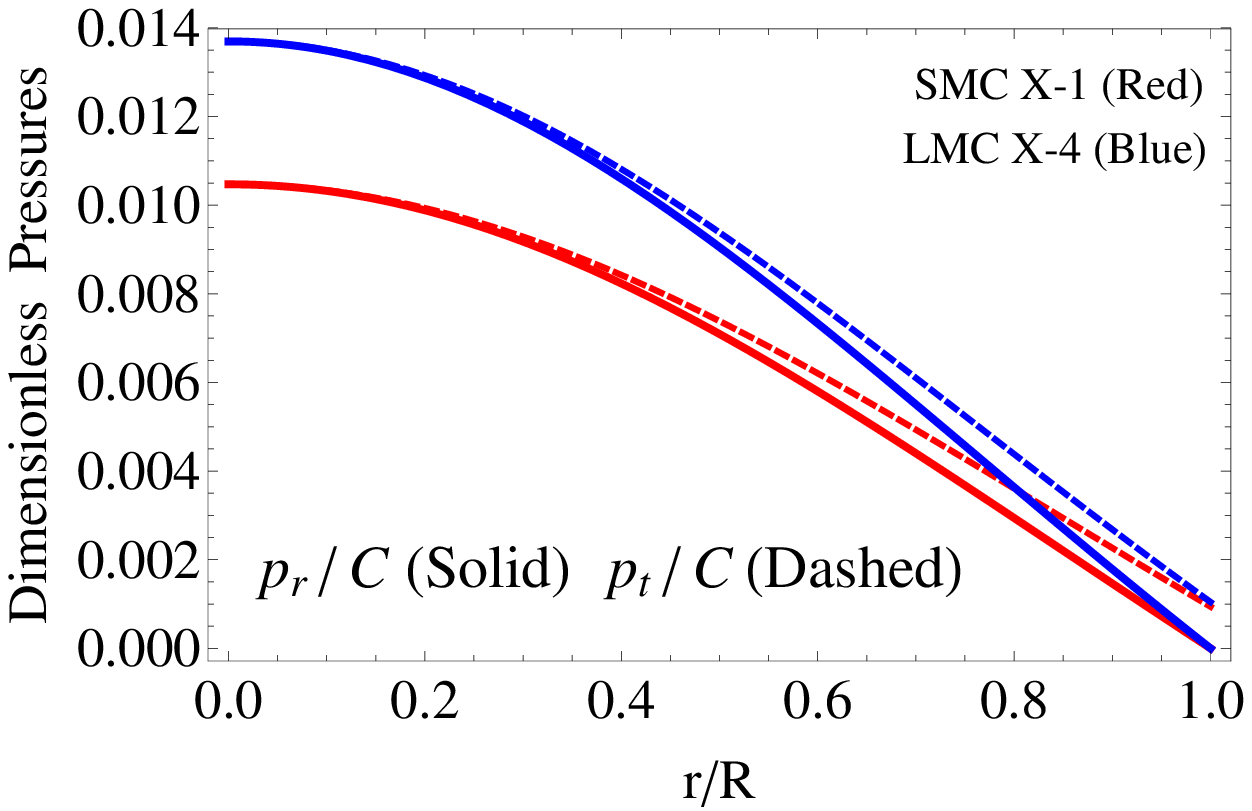}      
\includegraphics[width=0.34\textwidth]{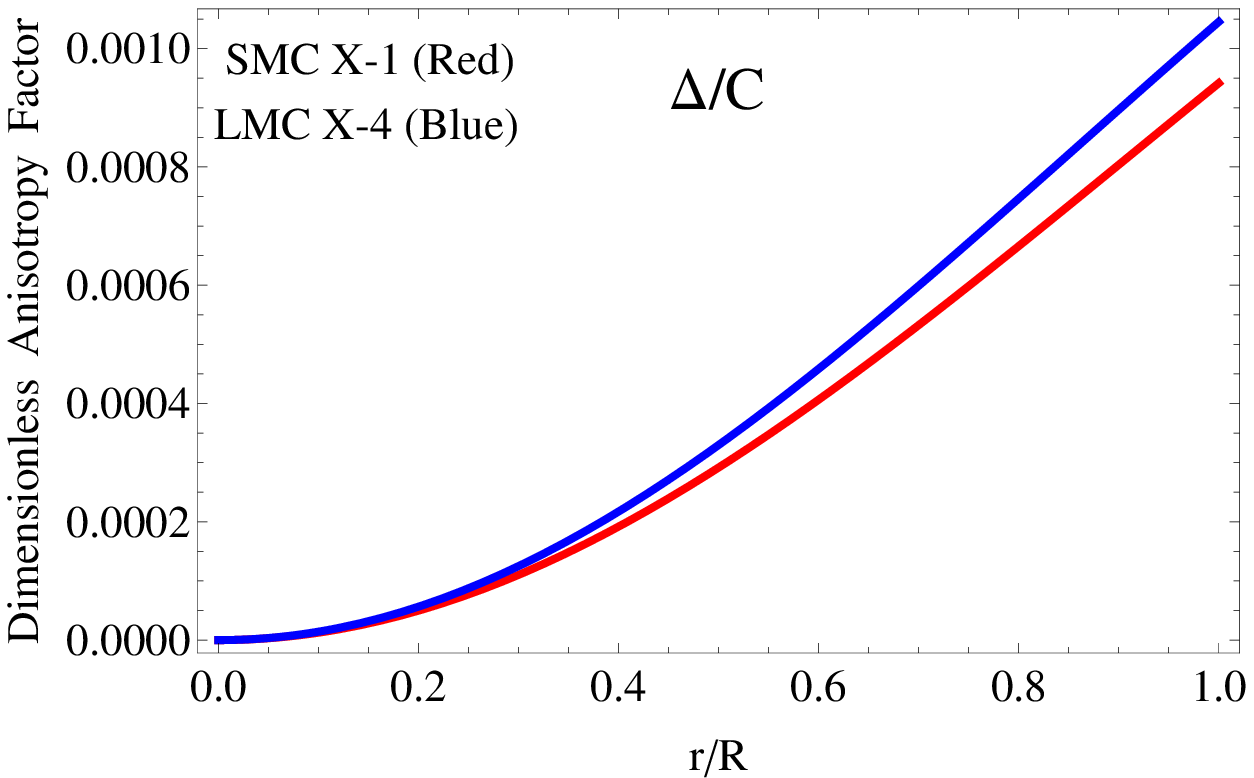}     \
\caption{Metric potential behaviours are shown in left panel of the upper row while the graphical behavior of density is shown in right panel of the upper row against radial coordinate $r/R$ by taking different values from table \ref{table1}. The left side of lower row shows the path of the radial and tangential pressures against the dimensionless radial variable $r/R$ while the right side shows the plot of anisotropy function against $r/R$.
}
\label{fig1}
\end{figure}

As the upper left panel in Fig. \ref{fig1} illustrates, both metric potential behaved as expected. It is important to note that we plotted the inverse of $e^{\lambda}$ in order to show that both functions match at the boundary. This confirms that the junction  procedure is well--posed. {Besides, in order to avoid undesirable physical behaviors the radial coordinate in Eq. (\ref{eq24}) should satisfy $r\neq \pm \sqrt{\frac{1}{C}\left(1-\frac{A^{2}}{\bar{B}^{2}}\right)}$ for all $r\in [0,R]$.} To build this panel we have used the numerical data listed in table \ref{table1}. In considering the thermodynamic observables $\{\rho,p_{r},p_{\perp}\}$, they all should satisfy some rules. First, all of them must have their maximum values attained at $r=0$. This means that these functions are monotonous decreasing with increasing $r$ inside the compact object. Second, {a positive defined behavior is required, in order to avoid a non--physical situation.} Furthermore, for the radial and tangential pressures is required: $p_{t}>p_{r} \Rightarrow \Delta>0$, assuring a healthy stellar interior. As said before, a positive anisotropy factor in principle enhances the stability and hydrostatic equilibrium of the configuration (see below for further details). Additionally, the radial pressure must vanish at the surface $p_{r}(R)=0$, determining the size of the object. {The previous conditions are subject to some restrictions imposed on the  
parameters $\{A,\bar{B},C\}$. At the center, the density $\rho$ and the radial pressure $p_{r}$ provide the following constraints}
\begin{eqnarray}\label{eq40}
p_{r}(0)&>&0 \Rightarrow \bar{B}>\frac{3\sqrt{2}}{10}A, \\ \label{eq41}
\frac{p_{r}(0)}{\rho(0)}&\leq&1\Rightarrow \bar{B}\leq \frac{3\sqrt{2}}{7}A, \\ \label{eq42}
\rho(0)&>&0\Rightarrow C>0.
\end{eqnarray}

So from Eqs. (\ref{eq40})--(\ref{eq41}) one has
\begin{equation}\label{eq43}
\frac{3\sqrt{2}}{10}A< \bar{B}\leq \frac{3\sqrt{2}}{7}A. 
\end{equation}
In table \ref{table2} we have placed the central values for density and radial pressure and the surface density of mentioned objects. The order of magnitude of these quantities is consistent with stars containing a core with strange matter (quark stars) \cite{r47,nature}. The right upper panel in Fig. \ref{fig1} shows the trend of the density $\rho$ inside the star. As can be seen, this observable is positive defined everywhere and has its maximum attained at the origin. The lower panel in the same figure display the behaviour of the pressures waves $p_{r}$ and $p_{\perp}$ and anisotropy factor $\Delta$. It is observed that the transverse pressure $p_{\perp}$ dominates the radial one $p_{r}$ inside the star, ensuring $\Delta>0$ everywhere. These features shows that the stellar interior is well--behaved. Notwithstanding, the matter distribution must satisfy  additional restrictions to confirm its viability in describing a stellar interior. Those are: i) positive and well behaved energy--momentum tensor and ii) preservation of causality condition. {
The former indicates that any well behaved energy--momentum tensor should meet the following inequalities known as energy conditions} \citep{e4,visserbook}

\begin{enumerate}
    \item Null energy condition (NEC): $\rho+p_{r} \geq 0$, $\rho+p_{t}+\frac{E^{2}}{4\pi} \geq 0$.
    \item Weak energy condition (WEC): $\rho+\frac{E^{2}}{8\pi} \geq 0$, $\rho+p_{r} \geq 0$, $\rho+p_{t}+\frac{E^{2}}{4\pi} \geq 0$ .  
    \item Strong energy condition (SEC): $\rho+p_{r} \geq 0$, $\rho+p_{t}+\frac{E^{2}}{4\pi} \geq 0$, 
    $\rho+2p_{t}+p_{r}+\frac{E^{2}}{4\pi} \geq 0$.
    \item Dominant energy condition (DEC): $\rho +\frac{E^{2}}{8\pi}-|p_{r}-\frac{E^{2}}{8\pi}|\geq0$, $\rho +\frac{E^{2}}{8\pi}-|p_{t}+\frac{E^{2}}{8\pi}|\geq0$.
    \item Trace energy condition (TEC): $\rho-p_{r}-2p_{t}
    \geq 0$.
\end{enumerate}

\begin{figure}[H]
\centering
\includegraphics[width=0.32\textwidth]{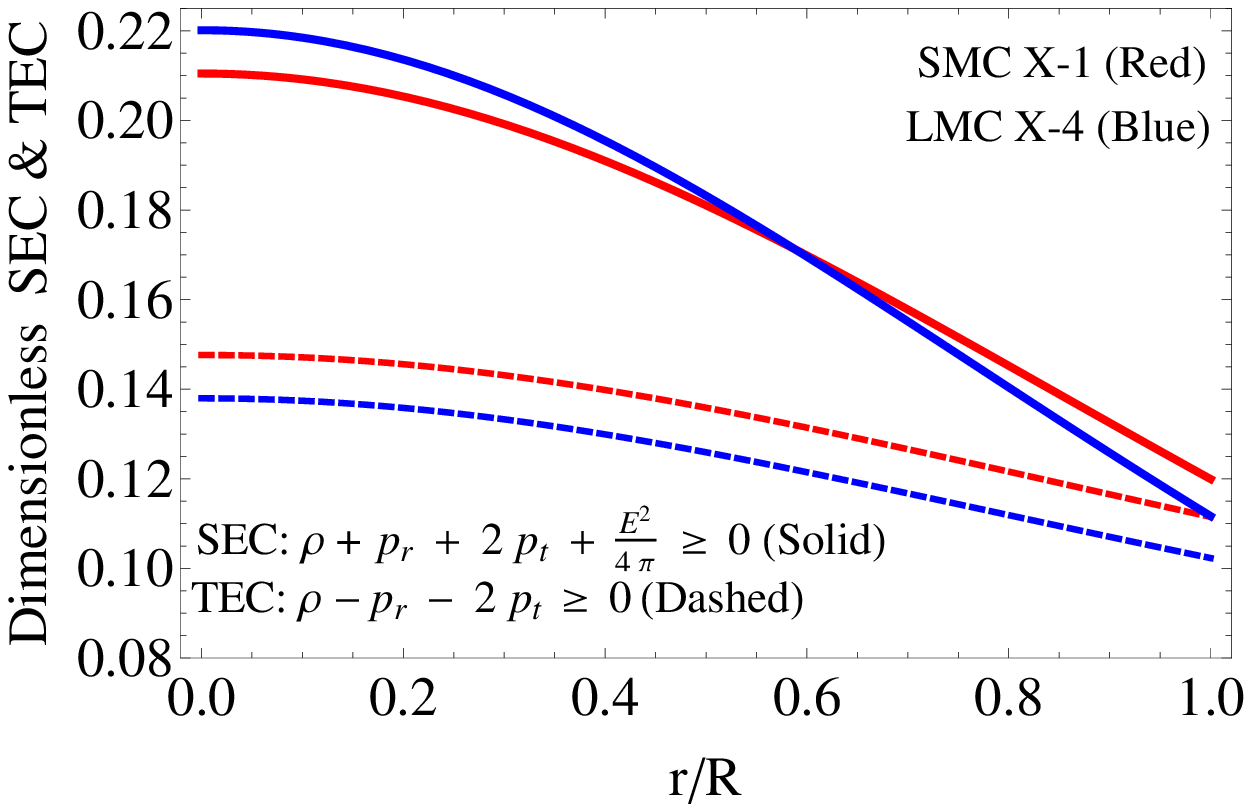}
\includegraphics[width=0.32\textwidth]{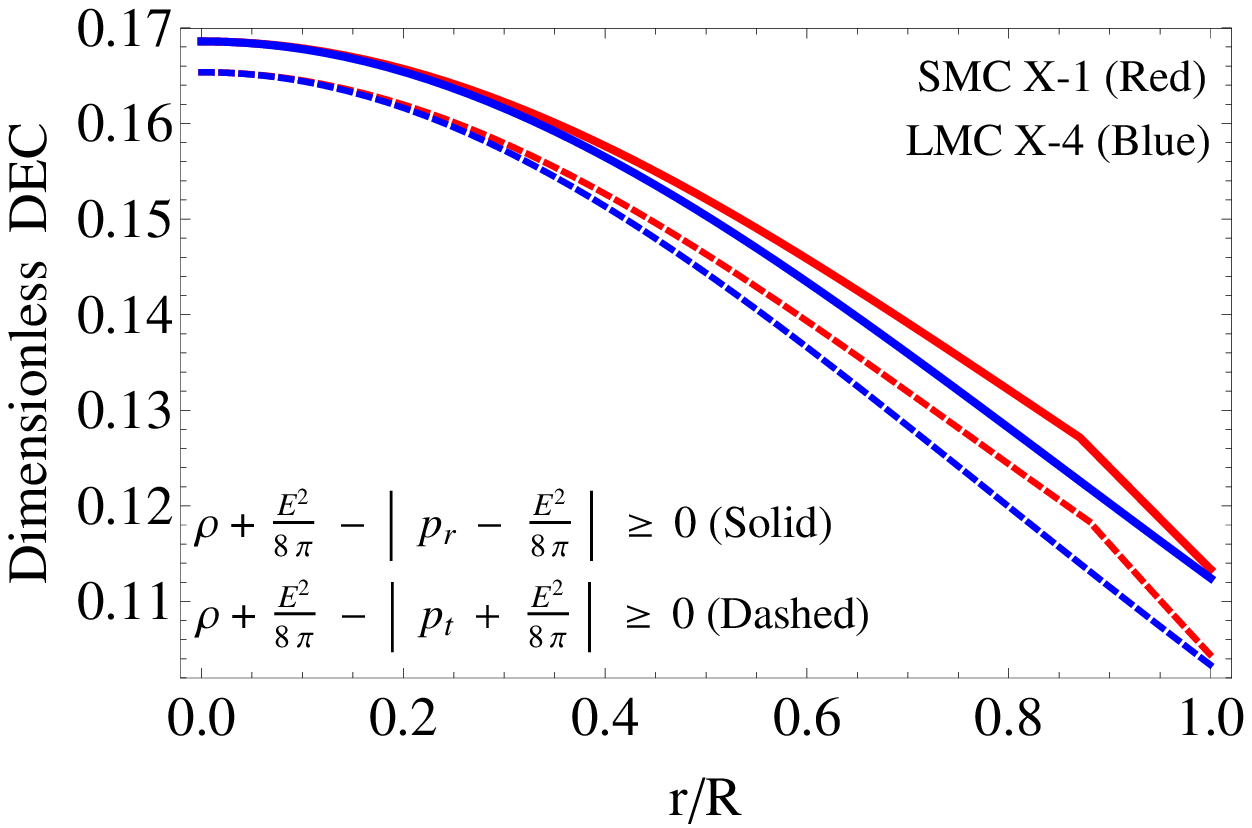}   
\caption{
Left panel of the figure shows the behavior of the SEC and TEC against the dimensionless radial coordinate $r/R$ while the right panel shows the plot of DEC versus radial coordinate $r/R$. The graphs has plotted by using different values mentioned in table \ref{table1}. 
}
\label{fig2}
\end{figure}

From Fig. \ref{fig2} (left  panel) is clear that SEC and TEC are satisfied everywhere, while the right panel exhibiting the DEC shows that this condition is also satisfied. Hence, the stellar interior is described by a well defined and positive energy--momentum tensor. It is worth mentioning that the NEC and WEC are contained by SEC and DEC.  

\begin{table}[H]
\caption{Numerical values for central and surface density, central pressure, critical adiabatic index and central adiabatic index for different values listed in table \ref{table1}.}
\label{table2}
\begin{tabular*}{\textwidth}{@{\extracolsep{\fill}}lrrrrrrrl@{}}
\hline
\multicolumn{1}{c} {Strange}&  
\multicolumn{1}{c} {$\rho(0)$}& 
\multicolumn{1}{c} {$\rho(R)$}& 
 \multicolumn{1}{c} {$p_{r}(0)$}& \multicolumn{1}{c}{$\Gamma_{\text{crit}}$} & \multicolumn{1}{c}{$\Gamma$}  \\
 ~~~~~~~\ Star&$\times 10^{15}\  [\text{g}/\text{cm}^{3}]$&$\times 10^{14}\  [\text{g}/\text{cm}^{3}]$&$\times 10^{34}\  [\text{dyne}/\text{cm}^{2}]$\\ 
\hline
SMC X--1 ~\ \cite{r47} & 1.07168&6.78682&5.64072&1.5003&2.8551 \\
\hline
LMC X--4 ~\ \cite{r47}& $1.15024$&$6.70688$& $7.91838$&$1.5280$ & $2.5726$    \\
\hline
\end{tabular*}
\end{table}

\subsection{Electric Properties}
As we are dealing with a charged configuration, electric properties should also meet some requirements. Both electric charge $q(r)$ and electric field $E(r)$ must be strictly positive and increasing functions with radius, meaning that at the origin both must be null \i.e, $q(0)=E(0)=0$. From expressions (\ref{eq27}) and (\ref{eq29}) the electric field is given by
\begin{equation}\label{eq44}
E(r)=\sqrt{\frac{C^2 r^2 \left[4 \bar{B} \left(C r^2-2\right)+3 A \sqrt{2-C r^2}\right]}{4\left[1+8\pi\chi\right]
\left[1+C r^2\right]^2 \left[\bar{B} \left(C r^2-2\right)+A \sqrt{2-C r^2}\right]}},
\end{equation}
{and from Eq. (\ref{eq5}) the electric charge is given by
\begin{equation}\label{eq45}
q(r)=r^{2}\, \sqrt{-F^{tr}F_{tr}}=r^{2}\, \sqrt{E^{2}(r)}=r^{2}E(r)=r^{2}\sqrt{\frac{C^2 r^2 \left[4 \bar{B} \left(C r^2-2\right)+3 A \sqrt{2-C r^2}\right]}{4\left[1+8\pi\chi\right]
\left[1+C r^2\right]^2 \left[\bar{B} \left(C r^2-2\right)+A \sqrt{2-C r^2}\right]}}.
\end{equation}}

\begin{figure}[H]
\centering
\includegraphics[width=0.32\textwidth]{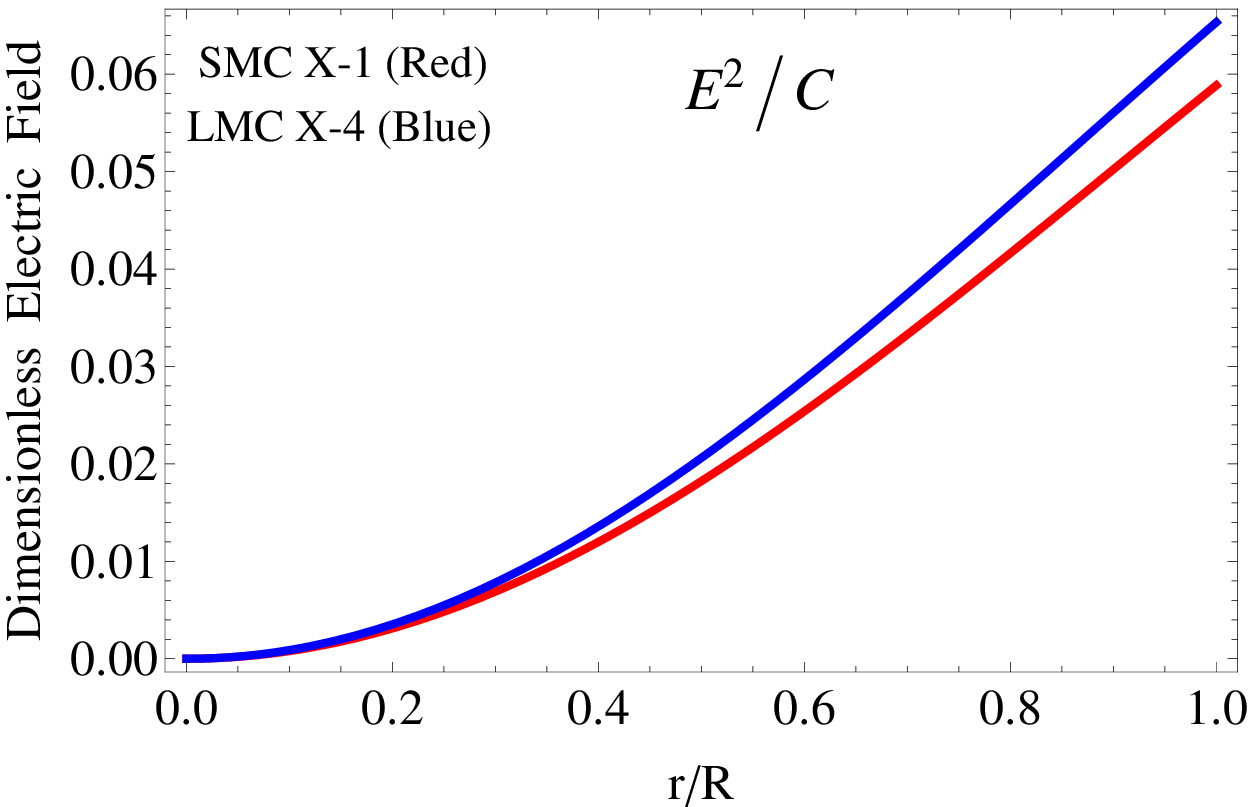}
\includegraphics[width=0.32\textwidth]{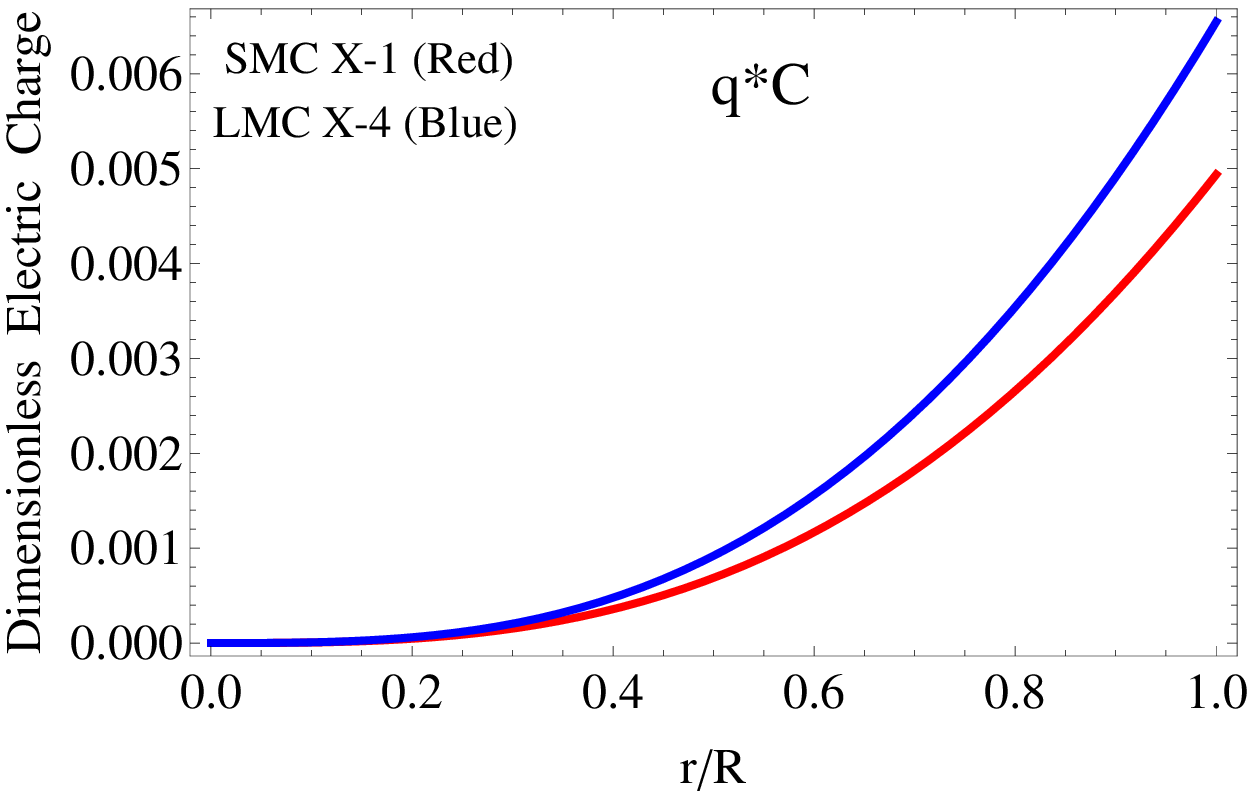}
\includegraphics[width=0.32\textwidth]{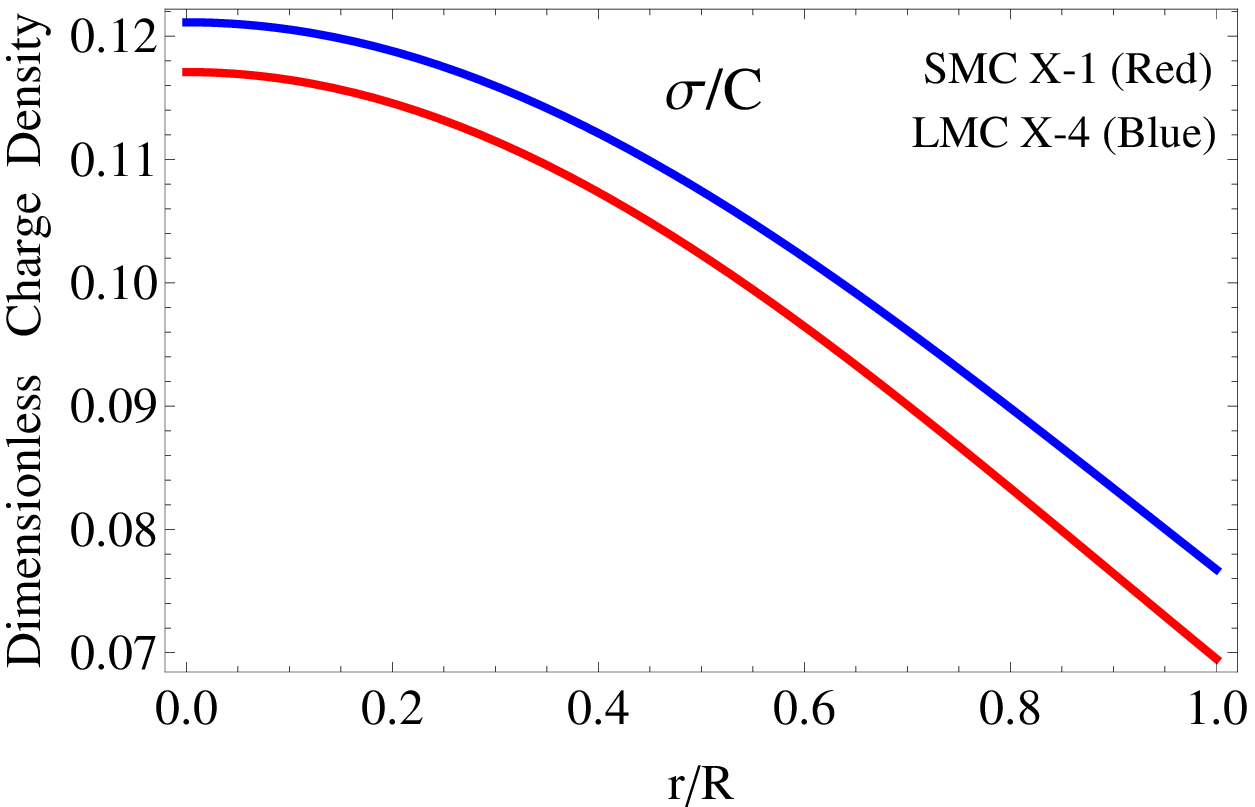}
\caption{
 The electric field trend against dimensionless radial coordinate $r/R$ is shown in the left panel. The trend of electric charge versus the dimensionless radial coordinate $r/R$ is shown in the middle panel while right side of the figure shows the graphical behavior of charge density versus $r/R$. These plots were built using different values mentioned at table \ref{table1}.
}
\label{fig3}
\end{figure}

From the above expressions is clear that at $r=0$ both quantities are zero, as can be confirmed in left and middle panels of Fig. \ref{fig3}. Additionally, we have checked the trend of charge density $\sigma$ inside the star. The general expression to obtain $\sigma$ is
\begin{equation}\label{eq46}
\sigma(r)=e^{-\lambda/2}\frac{\left( r^{2}E \right)^{\prime}}{4\pi r^{2}}.    
\end{equation}

The right panel of Fig. \ref{fig3} illustrates the behaviour of this quantity. Unlike the electric charge, it has its maximum attained at the center and decreases monotonously towards the boundary of the compact object. In table \ref{table3} are shown the values reached by $q$ and $E$ at the surface of the structure. As was point out in \cite{e6}, to appreciate any effect on the phenomenology of compact stars, the electric field have to be huge. Specifically, the order of magnitude of the electric field in the compact object should be $\sim 10^{21}\ [V/m]$, implying $Q\sim 10^{20}\ [C]$. As it is observed, the order of magnitude of these quantities is enough to appreciate the impact of electric components in the current scenario. Moreover, these values are within the upper limits reported in previous studies \cite{f1,f2,f3,f4}. To account the effects of parameter $\chi$ in electric properties, we have computed the numerical data presented in table \ref{table33} for different values of $\chi$. As can be seen, as $\chi$ decreases both $E(R)$ and $q(R)$ reach greater values.  

\begin{table}[H]
\caption{The surface gravitational red--shift, electric field and electric charge for values depicted in table \ref{table1}. }
\label{table3}
\begin{tabular*}{\textwidth}{@{\extracolsep{\fill}}lrrrrrrrl@{}}
\hline
Strange & \multicolumn{1}{c}{ Lower bound} & \multicolumn{1}{c}{$z_{s}$} &  \multicolumn{1}{c}{Upper bound}&
\multicolumn{1}{c}{$E(R)$} &
\multicolumn{1}{c}{$q(R)$}\\
~\ Star &$z_{s}$ & &$z_{s}$ &$\times  10^{21} [V/cm]$ & $\times 10^{20} [C] $\\
\hline
SMC X--1 ~\ \cite{r47} &0.013763 &0.259026&2.38478&2.18009&1.66914\\
\hline
LMC X--4 ~\ \cite{r47} &$0.018692$&$0.324997$& $2.55665$ &2.37971&1.82197   \\
\hline
\end{tabular*}
\end{table}

\begin{table}[H]
\caption{The surface gravitational red--shift, electric field and electric charge for values depicted in table \ref{table1} and different values of $\chi$. }
\label{table33}
\begin{tabular*}{\textwidth}{@{\extracolsep{\fill}}lrrrrrrrl@{}}
\hline
Strange &  \multicolumn{1}{c}{ $\chi$}&\multicolumn{1}{c}{ Lower bound} &  \multicolumn{1}{c}{Upper bound}&
\multicolumn{1}{c}{$E(R)$} &
\multicolumn{1}{c}{$q(R)$}\\
~\ Star &&$z_{s}$  &$z_{s}$ &$\times  10^{21} [V/cm]$ & $\times 10^{20} [C] $\\
\hline
SMC X--1 ~\ \cite{r47} &0.2 &0.002809&2.06939&0.992808&0.760123\\
\hline
SMC X--1 ~\ \cite{r47} &1.2 &0.000546&2.01318&0.438413&0.335661\\
\hline
LMC X--4 ~\ \cite{r47} &0.2&0.003818& 2.09531 &1.08717&0.832372   \\
\hline
LMC X--4 ~\ \cite{r47} &1.2&0.000742& 2.01795 &0.480407&0.367814   \\
\hline
\end{tabular*}
\end{table}

\section{Hydrostatic Equilibrium and Stability}\label{sec5}

\begin{figure}[H]
\centering
\includegraphics[width=0.32\textwidth]{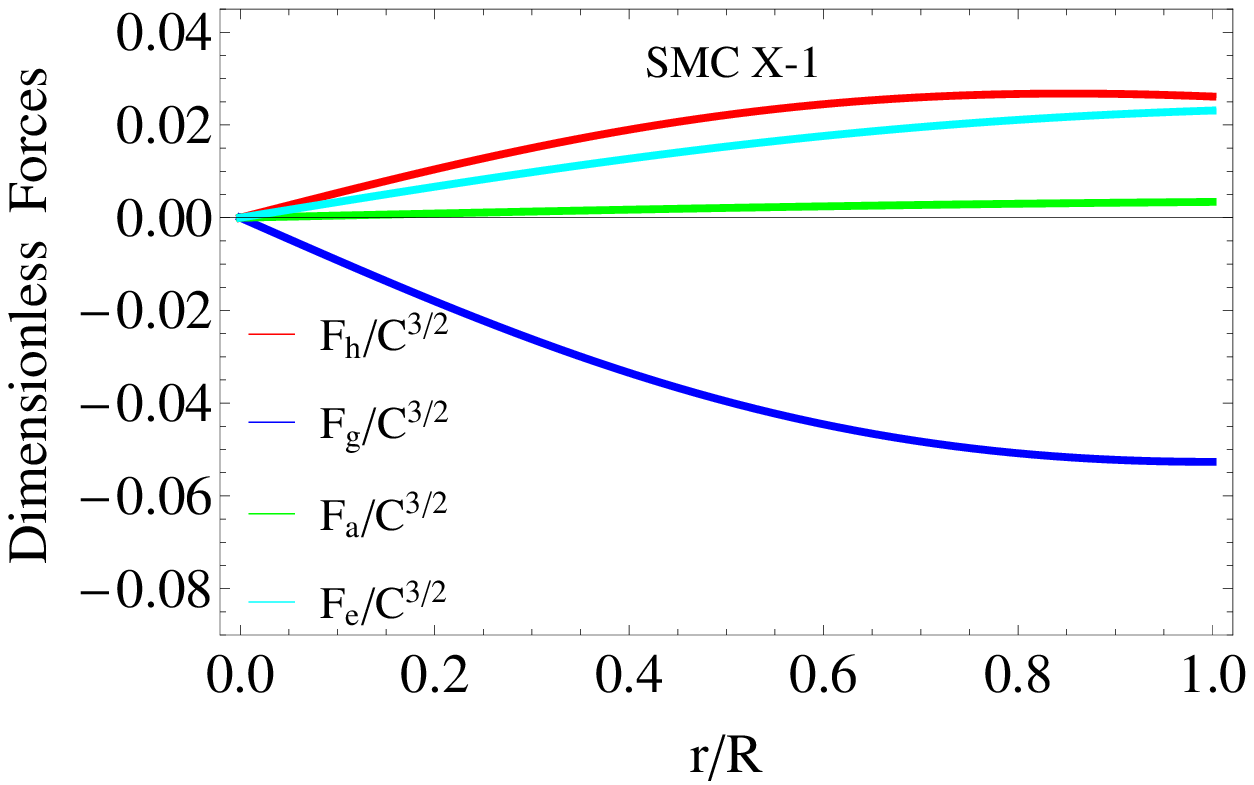}
\includegraphics[width=0.32\textwidth]{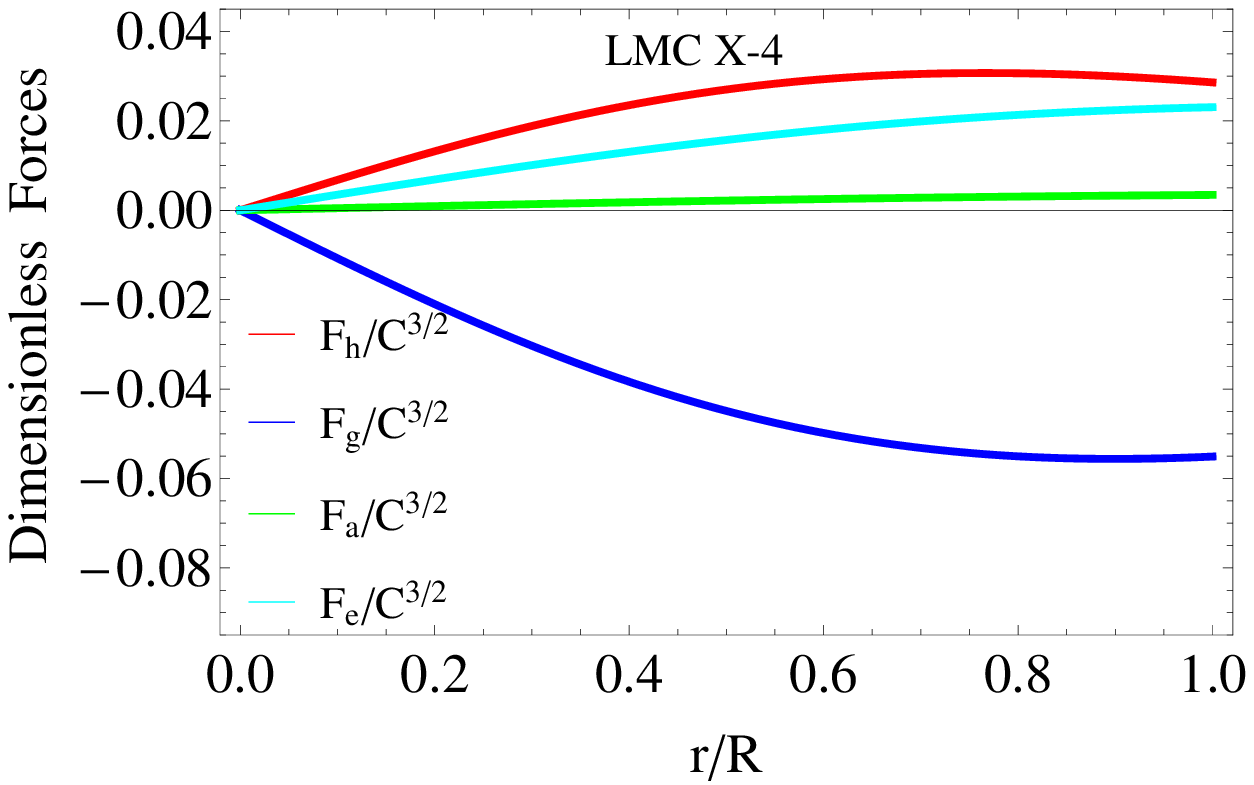}    
\caption{
The trend of the hydrostatic $F_{h}$, gravitational $F_{g}$, anisotropic $F_{a}$ and electric force gradients for each strange star candidate are shown in both panels of the figure. These plots were built for values mentioned in table \ref{table1}.
}
\label{fig4}
\end{figure}

For the isotropic scenario, the hydrostatic equilibrium of the structure is subject to the gravitational and hydrostatic gradients. However, in presence of extra ingredients, such as anisotropies and electric charge, the hydrostatic balance changes. Now, the configuration is under the action of gravitational $F_{g}$, hydrostatic $F_{h}$, anisotropy $F_{a}$ and electric $F_{e}$ gradients. If the structure is in equilibrium, all gradients must satisfy
\begin{equation}\label{eq47}
F_{h}+F_{a}+F_{g}+F_{e}=0,    
\end{equation}
or, in terms of the main physical variables
\begin{equation}\label{eq48}
\begin{split}
-p^{\prime}_{r}-\frac{1}{2}\,\eta^\prime \left({\rho}+{p}\right)+\frac{2}{r}\left(p_{\perp}-p_{r}\right)+\sigma \,E\,e^{\lambda/2}=0.
\end{split}
\end{equation}
Eq. (\ref{eq48}) is just the conservation equation (\ref{eq10}). It should be noted that when $\Delta\equiv p_{t}-p_{r}=0$ and $E=0$, {the original Tolman--Oppenheimer--Volkoff equation \cite{tolman,oppen} used to study the hydrostatic balance of compact perfect fluids, is regained.} The new pieces add into the system positive gradients ($F_{a}$ and $F_{e}$) \i.e, repulsive in nature. It is worth mentioning that the electric gradient is repulsive due to the Coulomb electric repulsion. These repulsive gradients help to offset the gravitational one. {The presence of these new components prevents the contraction of the object by gravitational action}. In Fig. \ref{fig4} is depicted {the action of the mentioned gradient at all points within the structure. As
can be seen, the configuration is in hydrostatic equilibrium.} Is remarkable to note that the electric component overcome the anisotropy one. This situation is connected with the inverse behavior that $\Delta$ and $E$ have with respect to the magnitude of parameter $\chi$. In this opportunity we have taken a small $\chi$, concretely $0.008$. This is so because the anisotropy gradient represented by the green curve has less contribution than the electric gradient (cyan curve) in the balance of the compact object. However the most important thing here is that both gradients in accordance with the hydrostatic gradient produce the balance against gravitational attraction. 

{Next, we analyze if the hydrostatic balance is stable or unstable. To do so, the following criteria are considered: i) the relativistic adiabatic index $\Gamma$, ii) the Harrison–-Zeldovich–-Novikov \cite{harrison,nikolov} condition and iii) the criterion based on sound speed of the matter distribution \cite{a19}. }

\begin{figure}[H]
\centering
\includegraphics[width=0.32\textwidth]{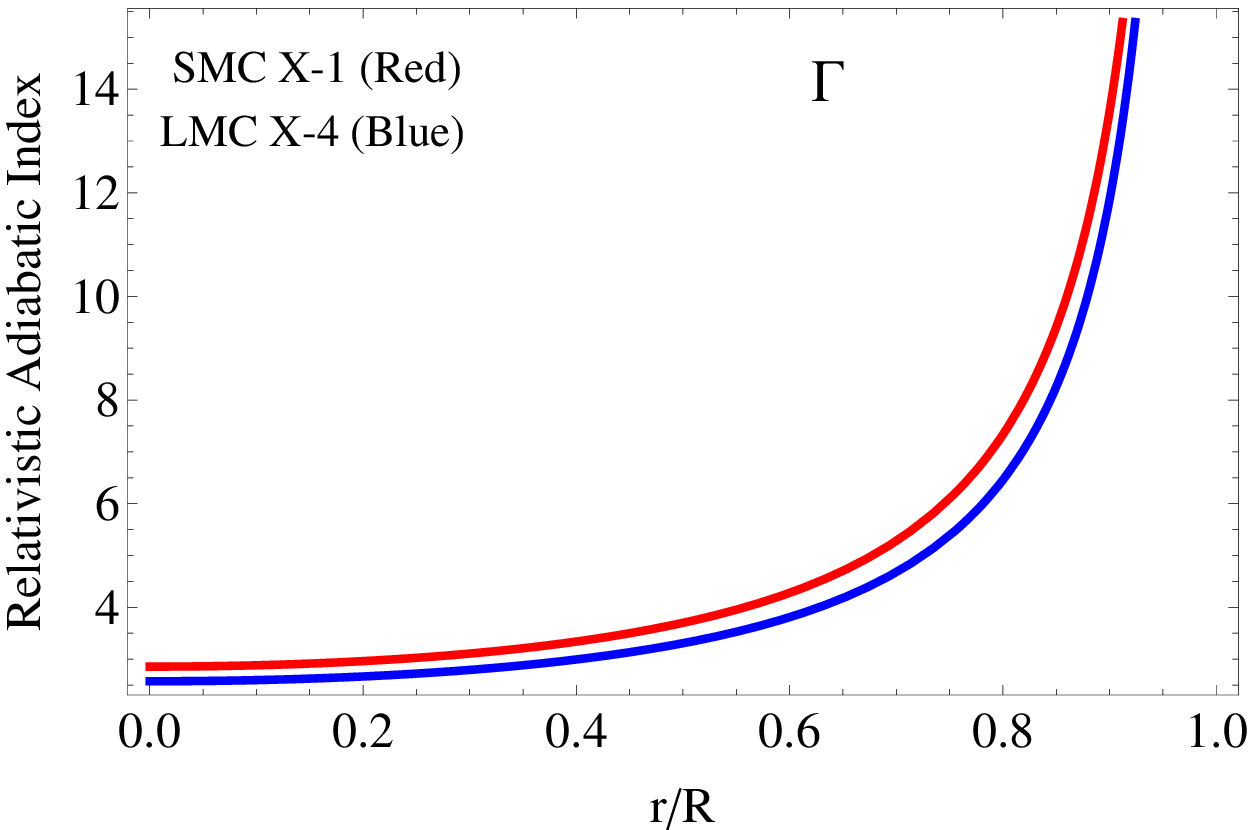}
\includegraphics[width=0.32\textwidth]{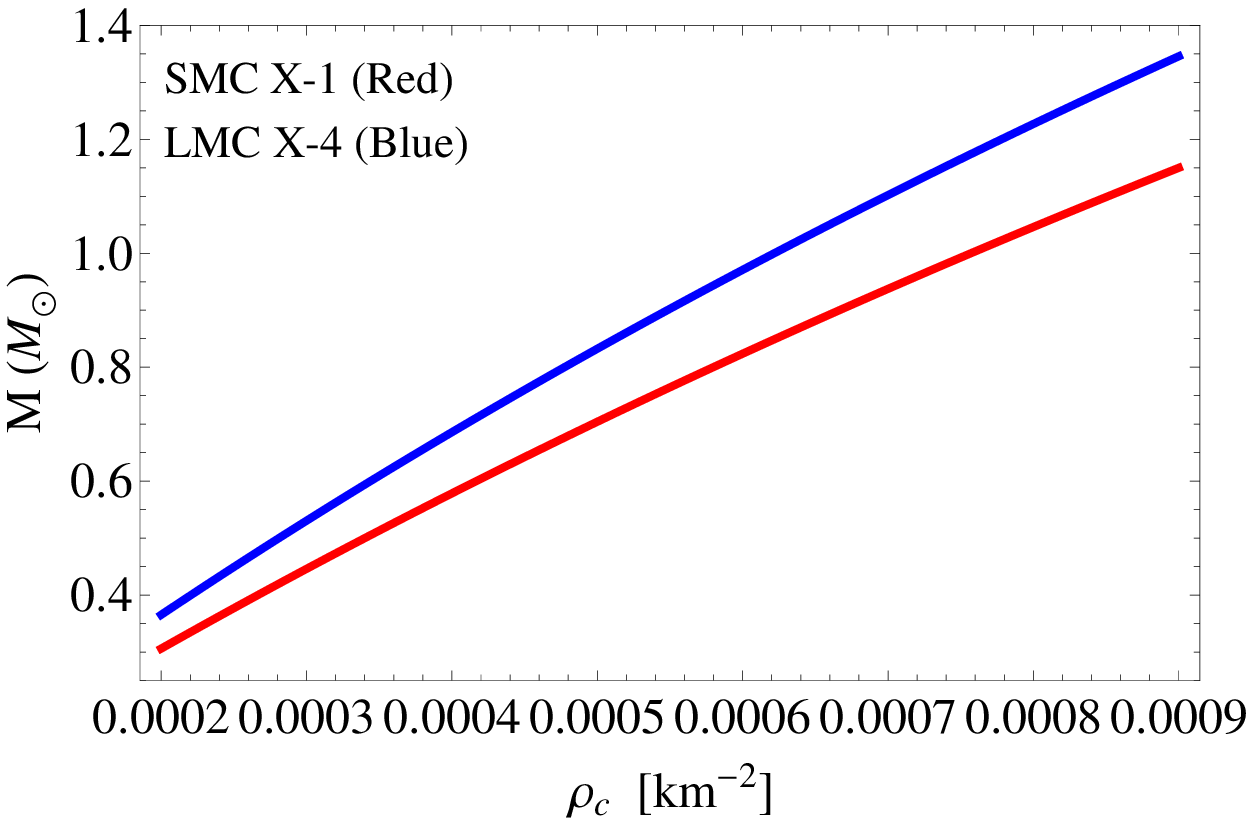}   
\caption{
 The relativistic adiabatic index path against the dimensionless radial coordinate $r/R$ is shown in left panel, whereas the graphical behavior of mass function versus $r/R$ is shown in right panel of the figure for different values mentioned in previous tables.
}
\label{fig5}
\end{figure}

{As it is well known, in the classical domain (Newtonian fluid) taking into account a perfect fluid content, the stability condition is $\Gamma>4/3$ \cite{a1,a8}. Notwithstanding, in contrast with the relativistic regime subject to an imperfect matter distribution, the scene changes radically. In this concern  local imperfections (or anisotropies) within the material content insert harsh modifications for the stability condition \cite{a9,a10}. Under this situation, the stability condition becomes}
\begin{equation}\label{eq49}
\Gamma>\frac{4}{3}+\left[\frac{1}{3}\,\kappa\,\frac{\rho_{0}\,p_{r0}}{|p^{\prime}_{r0}|}r+\frac{4}{3}\frac{\left(p_{\perp 0}-p_{r0}\right)}{|p^{\prime}_{r0}|r}\right]_{max},
\end{equation}
{where $\left[\rho_{0}, p_{r0}, p_{t0}\right]$ are the initial values of the thermodynamic observables, when the matter distribution is in static equilibrium. The terms enclosed in the brackets are representing the relativistic adjustment and the point anisotropies contributions. It should be noted that local anisotropies can be seen as a stabilizer mechanism. However, as Chandrasekhar warned \cite{chandra1,chandra2}, these relativistic corrections to $\Gamma$ could insert some instabilities within the matter distribution. To overcome this problem, new constraints on $\Gamma$ were imposed \cite{mousta}. Specifically, a critical adiabatic index  $\Gamma_{\text{crit}}$ depending on the critical value of the amplitude of the Lagrangian displacement from equilibrium and the compactness factor $u\equiv M/R$ were fixed. Particularly, this constraint reads as}
\begin{equation}\label{eq50}
\Gamma_{\text{crit}}=\frac{4}{3}+\frac{19}{21}\,u,    
\end{equation}
being the stability condition $\Gamma\geq \Gamma_{\text{crit}}$,
where $\Gamma$ is computed from \cite{a7}
\begin{equation}\label{eq51}
\Gamma=\frac{\rho+p_{r}}{p_{r}}\frac{d\,p_{r}}{d\,\rho}.    
\end{equation}            
From the left panel of Fig. \ref{fig5} it is appreciated that in all cases the relativistic adiabatic index $\Gamma$ is greater than $4/3$. Commonly, it is assumed that when $\Gamma>4/3$ at $r=0$ the system is stable under this criterion. {However, as pointed out before, a more precise analysis indicates} $\Gamma\geq\Gamma_{\text{crit}}$ at $r=0$. The fifth and sixth columns of table \ref{table2} presents the corresponding values for $\Gamma$ and $\Gamma_{\text{crit}}$ for each considered compact structure. The Harrison et al. \cite{harrison} and Zeldovich–-Novikov \cite{nikolov} methods propose that any fluid configuration is stable if the mass is a increasing function with respect to the central density $\rho(0)=\rho_{c}$ \i.e, $\frac{\partial M(\rho_{c})}{\partial \rho_{c}}>0$, otherwise the model is unstable. 
For the present model, the mass as a function of the central density has the following form
\begin{equation}\label{eq52}
M(\rho_{c})=\frac{1}{8} R \left(6-\frac{6}{1+\frac{16}{9} \pi \rho_{c} R^2}+\frac{\left[\frac{16}{9} \pi \rho_{c}\right]^2
R^4 \left[4 \bar{B} \left(\frac{16}{9} \pi \rho_{c}R^2-2\right)+3 A \sqrt{2-\frac{16}{9} \pi \rho_{c} R^2}\right]}{\left[1+\frac{16}{9} \pi \rho_{c} R^2\right]^2 \left[\bar{B} \left(\frac{16}{9} \pi \rho_{c}R^2-2\right)+A \sqrt{2-\frac{16}{9} \pi \rho_{c}R^2}\right]
\left[1+8 \pi  \chi \right]}\right).    
\end{equation}

Right panel of Fig. \ref{fig5} is illustrates the trend of total mass against the central density. As can be seen the criteria is satisfied, thus the model is stable under this criteria.

\begin{figure}[H]
\centering
\includegraphics[width=0.32\textwidth]{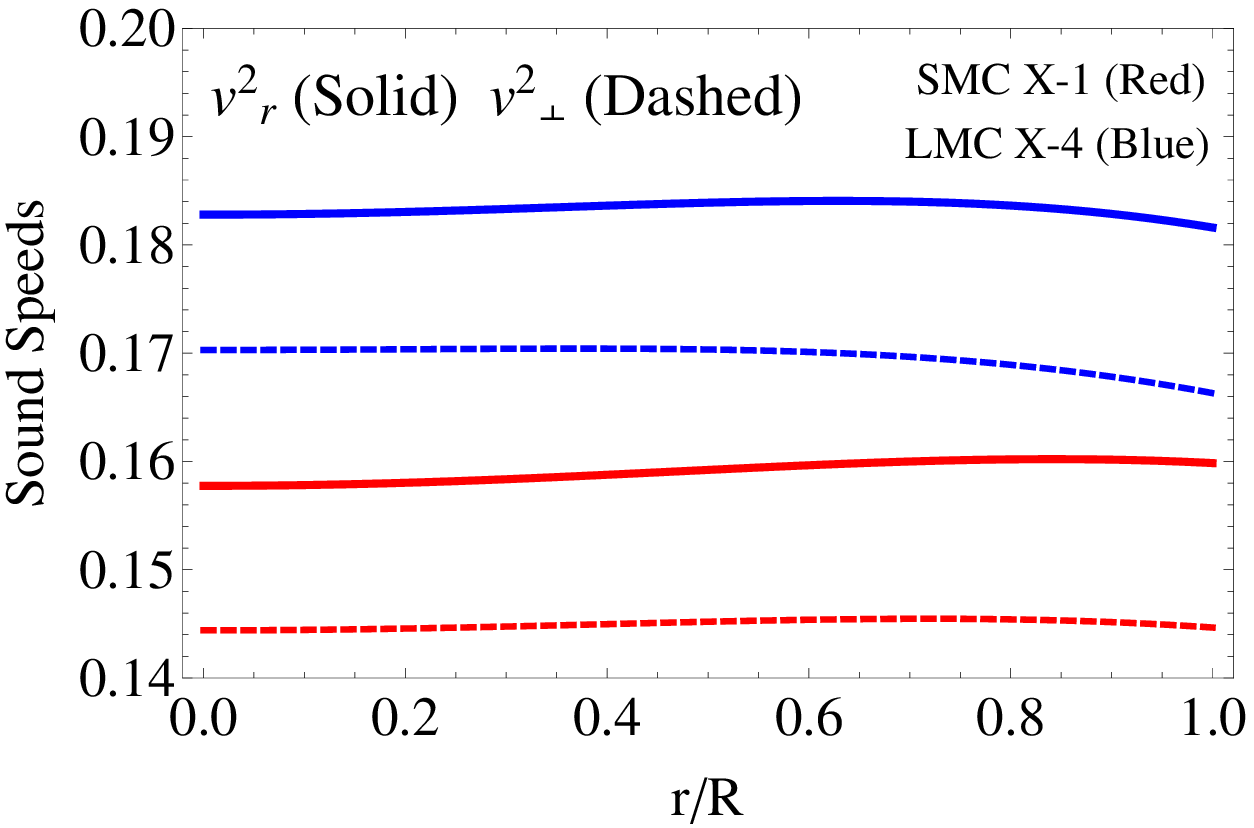}
\includegraphics[width=0.33\textwidth]{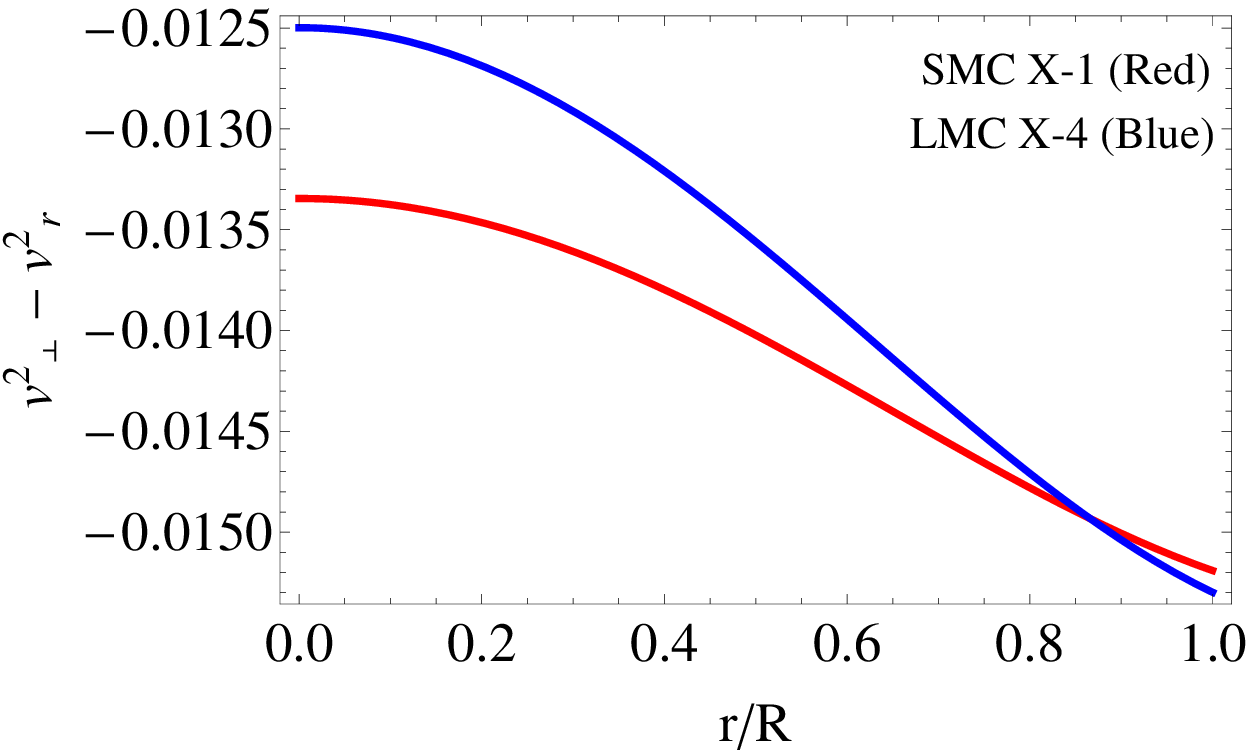}    
\includegraphics[width=0.33\textwidth]{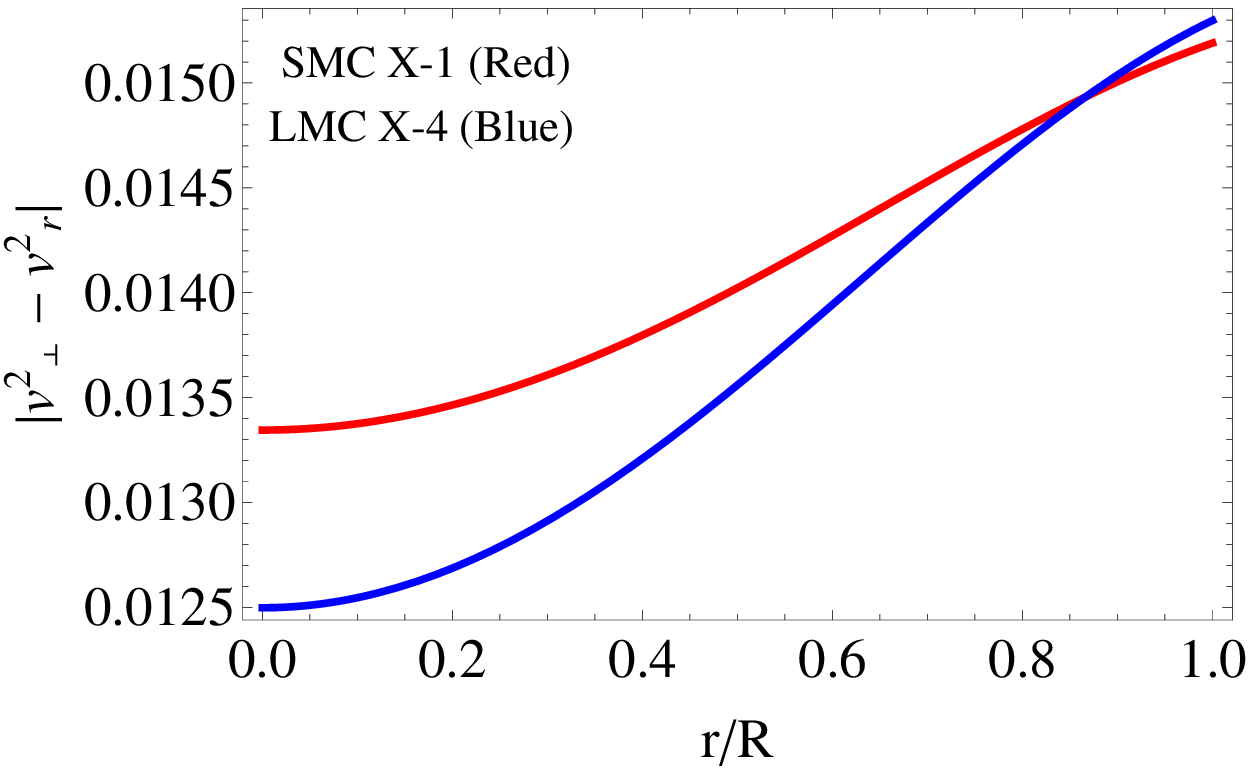}      

\caption{The path of the radial and transverse speed of sounds against the dimensionless quantity $r/R$ is shown in left panel of the figure. The difference of square of sound speed velocities and its absolute value versus $r/R$ are shown in middle and right panels. These plots were built for different values mentioned in table \ref{table1}. }
\label{fig6}
\end{figure}

Finally, we check the stability of the system by means of the subliminal sound speeds of the pressure waves. Before to proceed with the criteria it is important to see if the matter distribution is fulfilling the causality condition 
\begin{equation}\label{eq53}
0\leq v^{2}_{r}=\frac{dp_{r}}{d\rho}\leq 1 \quad \mbox{and} \quad  0\leq v^{2}_{\perp}=\frac{dp_{\perp}}{d\rho}\leq 1.  
\end{equation}
Eq. (\ref{eq53}) states that no signal can travel faster than the speed of light $c=1$. From left panel in Fig. \ref{fig6} we check that causality condition is preserved along the principal directions of the fluid sphere. Next, based on this important fact, the stable/unstable regions within the stellar, interior, when local anisotropies are there can be found as \cite{a19}:
\begin{equation}\label{eq54}
\frac{\delta{\Delta}}{\delta{\rho}}\sim \frac{\delta\left({p}_{\perp}-{p}_{r}\right)}{\delta{\rho}} \sim \frac{\delta{p}_{\perp}}{\delta{\rho}}-\frac{\delta{p}_{r}}{\delta{\rho}}\sim v^{2}_{\perp}-v^{2}_r.    
\end{equation}
Taking into account equation (\ref{eq54}) one gets  $0\leq |v^{2}_{\perp}-v^{2}_{r}|\leq 1$ or equivalently
\begin{equation}
\begin{split}
     \label{eq55}
   -1\leq v^{2}_{\perp}-v^{2}_{r}\leq 1  = \left\{
	       \begin{array}{ll}
		   -1\leq v^{2}_{\perp}-v^{2}_{r}\leq 0~~ & \mathrm{Potentially\ stable\ }  \\
		 0< v^{2}_{\perp}-v^{2}_{r}\leq 1 ~~ & \mathrm{Potentially\ unstable}
	       \end{array}
	        \right\}.
	        \end{split}
	    \end{equation}
{Hence, the compact object is stable under radial perturbation if and only if the radial sound speed $v^{2}_{r}$ dominates at all points the transverse sound speed $v^{2}_{\perp}$. In the present case, it is evident that the difference of the square sound speeds is bounded between -1 and 0 and the absolute value of this quantity is between 0 and 1, as showed by middle and right panel of Fig.\ref{fig6}. Thus, all the regions inside the collapse structure are stable under this criterion.} 

\section{Astrophysical Observables }\label{sec6}

Since it is not possible to directly measure the properties of a stellar object,
from astrophysical techniques one can obtain a relevant quantity: the surface gravitational red--shift $z_{s}$, to  infer important properties such as the total mass $M$ and the radius $R$ of the star, its chemical composition, etc. In considering the total mass of the compact structure, when the matter distribution contains a charge component it increases by a certain amount, which is provided by the electric field. Integrating the field equation (\ref{eq6}) one gets
\cite{e2,e3,e8}
\begin{equation}\label{eqeq56}
e^{-\lambda\left(r\right)}=1-\frac{1}{r}\int^{r}_{0}  \left[8\,\pi\, \rho\left(x\right) \,x^{2}+E^{2}\left(x\right)\right]\,dx,
\end{equation}
this expression must coincide with exterior space--time at the boundary $\Sigma$, yielding to
\begin{equation}\label{eq57}
1-2\,\frac{M}{R}+\frac{Q^{2}}{R^{2}}= 1-\frac{1}{r}\int^{r}_{0}  \left[8\,\pi \,\rho\left(x\right)\, x^{2}+E^{2}\left(x\right)\right]\,dx,
\end{equation}
being $Q=q\left(R\right)$ and $M$ the total mass of the object. So, from (\ref{eq57}) it is possible to obtain
\begin{equation}\label{eq58}
M=\frac{1}{2}\int^{R}_{0}  \left[8\,\pi\, \rho\left(r\right)\, r^{2}+E^{2}\left(r\right)\right]\,dr +\frac{1}{2}\,\frac{Q^{2}}{R}.
\end{equation}
This total mass $M$ correspond to the gravitational mass evaluated at the surface $m_{g}\left(r\right)=M$. So, from (\ref{eq57}) it is clear that the gravitational function mass is given by,
\begin{equation}\label{eq59}
m_{g}\left(r\right)=m_{i}\left(r\right)+\frac{1}{2}\int_0^r{\frac{q^2(x)}{x^2}\,dx}+\frac{1}{2}\,\frac{q^2}{r}.
\end{equation}

Here, $m_{i}\left(r\right)$ is the usual definition of the mass within a radius $r$,
\begin{equation}\label{eq60}
m_{i}\left(r\right)=4\,\pi\int_0^r{{\rho(x)}\,x^2\,dx}. 
\end{equation}
{In the above expression we have denoted $m_{i}$ as the mass containing both rest and internal energy. This prescription allows us to distinguish between  $m_{i}\left(r\right)$ from the gravitational
mass $m_{g}\left(r\right)$ \cite{e8}.} On the other hand, associated with gravitational mass is the so--called compactness factor or mass--radius ratios $u\equiv M/R$. For this particular model the gravitational mass functions and compactness factor are given by
\begin{equation}\label{eq61}
m_{g}\left(r\right)=\frac{r}{8}  \left[6-\frac{6}{1+C\, r^2}+\frac{C^2\, r^4 \left(4\, \bar{B} \left(C \,r^2-2\right)+3 \,A \,\sqrt{2-C \, r^2}\right)}{\left(1+C\,
r^2\right)^2 \left(\bar{B}\, \left(C\, r^2-2\right)+A \,\sqrt{2-C \,r^2}\right) \left(1+8 \,\pi \, \alpha \right)}\right],
\end{equation}
\begin{equation}\label{eq62}
u\left(r\right)=\frac{1}{8}  \left[6-\frac{6}{1+C\, r^2}+\frac{C^2\, r^4 \left(4\, \bar{B} \left(C \,r^2-2\right)+3 \,A \,\sqrt{2-C \, r^2}\right)}{\left(1+C\,
r^2\right)^2 \left(\bar{B}\, \left(C\, r^2-2\right)+A \,\sqrt{2-C \,r^2}\right) \left(1+8 \,\pi \, \alpha \right)}\right].
\end{equation}

As can be seen the electric component modified the mass and mass--radius ratio. This implies that the surface gravitational red--shift $z_{s}$ is also altered. In fact, $z_{s}$ depends on $u$ in the following way
\begin{equation}\label{eq63}
z_{s}=\frac{1}{\sqrt{1-2 \,u}}-1.    
\end{equation}

\begin{figure}[H]
\centering
\includegraphics[width=0.32\textwidth]{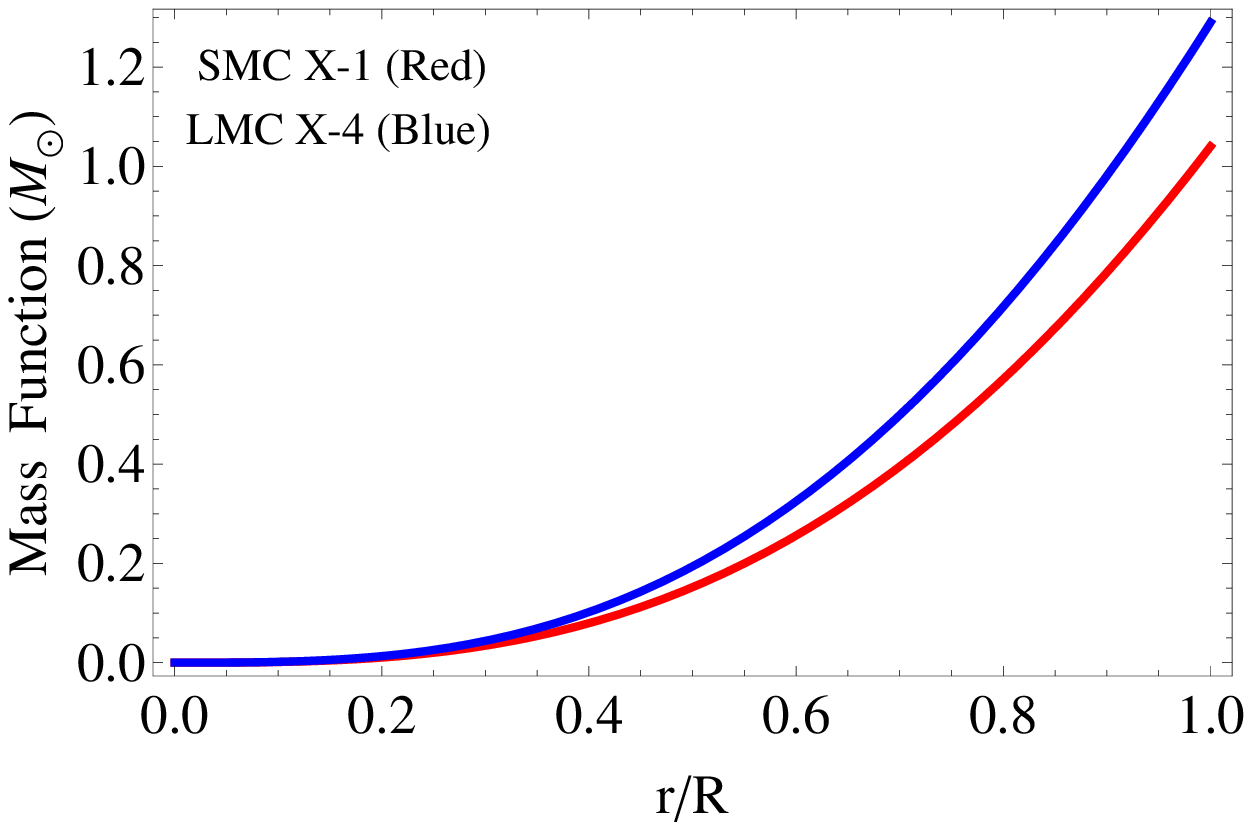}
\includegraphics[width=0.32\textwidth]{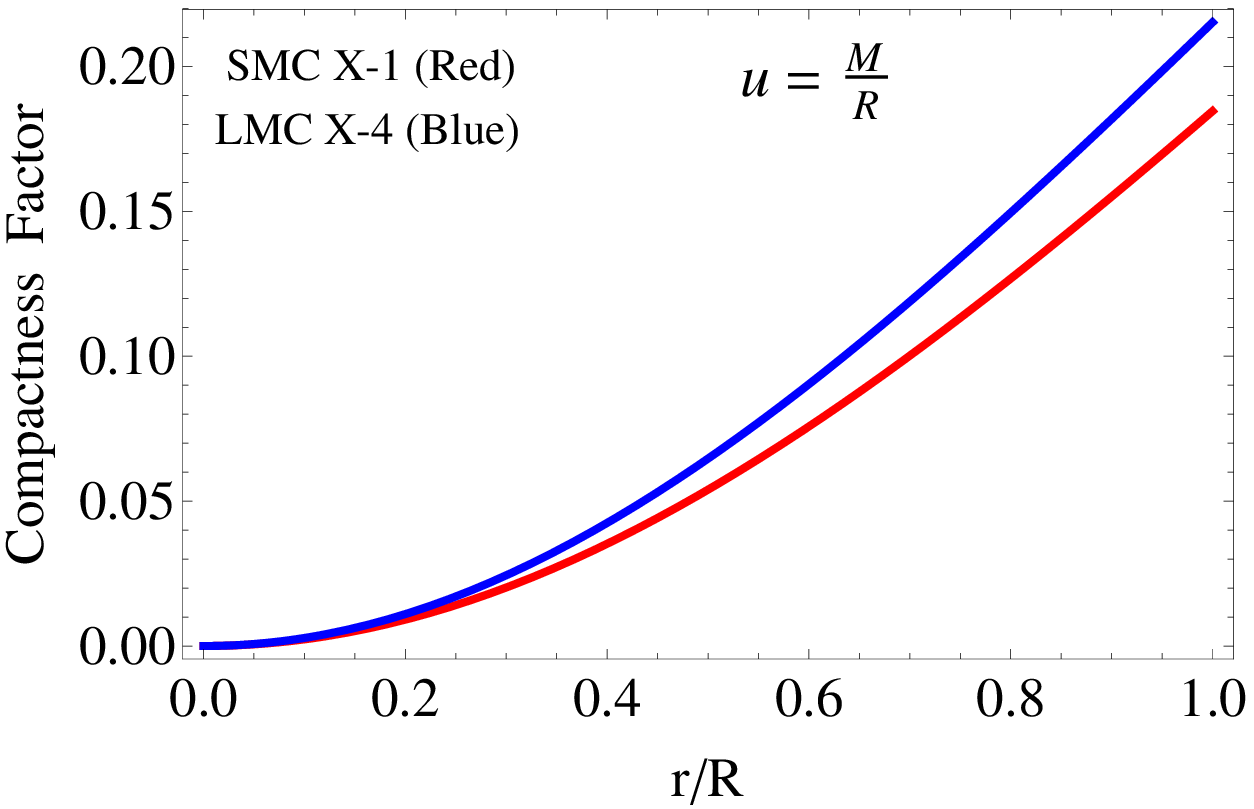}
\includegraphics[width=0.32\textwidth]{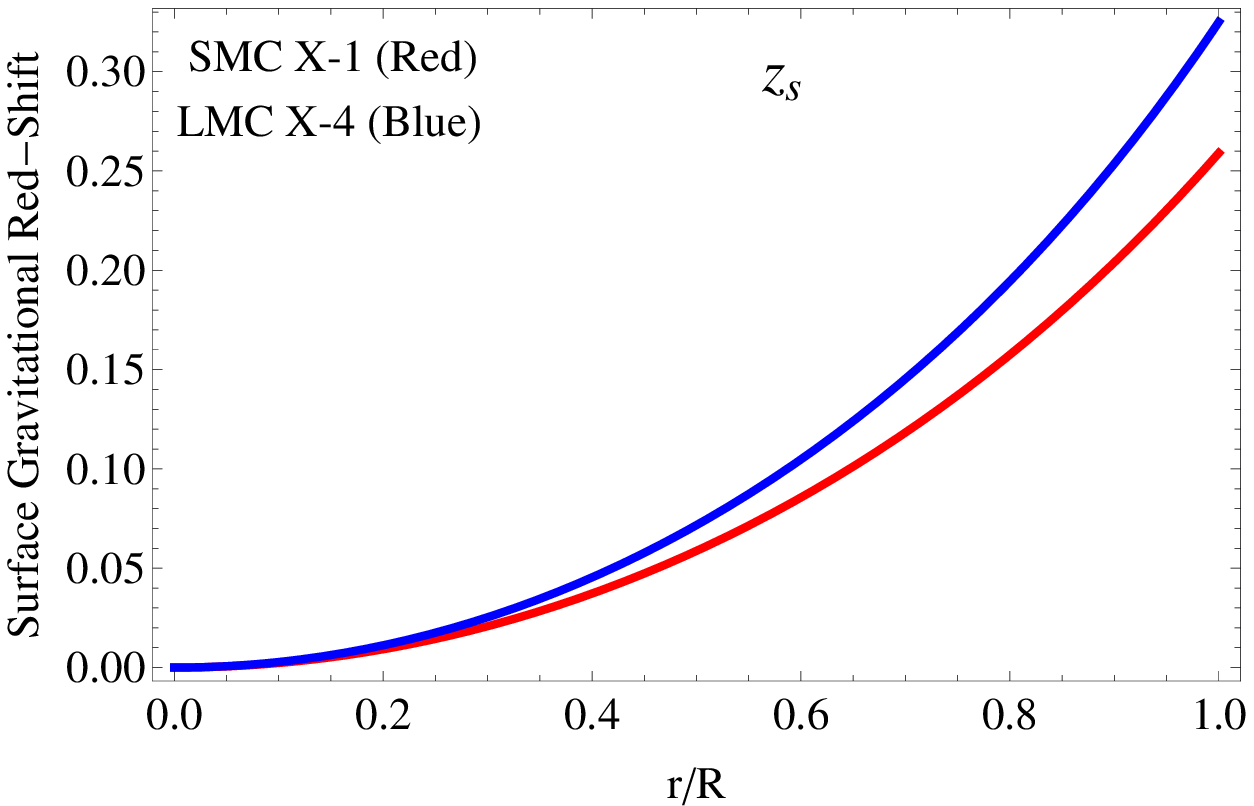}
\caption{
{\bf Left panel}: The mass function trend against the dimensionless radial coordinate $r/R$. {\bf Middle panel}: The compactness factor versus the dimensionless radial coordinate $r/R$. {\bf Right panel}: the gravitational surface red--shift versus $r/R$. These plots were built for different values mentioned in tables \ref{table1}.
}
\label{fig7}
\end{figure}

In Fig. \ref{fig7} the trend of (\ref{eq61})--(\ref{eq63}) is displayed for numerical values given in table \ref{table1} for different real compact objects. Furthermore, in tables \ref{table3} and \ref{table4} are exhibited the vales for $z_{s}$ and $u$ corresponding to these stars. Nevertheless, the electric field modifies the mass--radius relation, in such a way that the well--known Buchdahl limit $u=M/R=4/9$ \cite{buch} for isotropic uncharged fluid spheres can be overcome. This means that the upper bound exceeds the value $u=0.\bar{4}$ \cite{e9}, and also acquires a lower bound \cite{e7}
\begin{equation}\label{eq64}
\frac{Q^{2}\left(18R^{2}+Q^{2}\right)}{2R^{2}\left(12R^{2}+Q^{2}\right)}\leq \frac{M}{R}\leq \frac{2R^{2}+3Q^{2}+2R\sqrt{R^{2}+3Q^{2}}}{9R^{2}}.    
\end{equation}
As it is appreciated from table \ref{table4} the upper bound of $u$ surpasses the Buchdahl bound. For both lower and upper limits we have computed the corresponding $z_{s}$ shown by table \ref{table3}. In this respect, Ivanov \cite{iva}  pointed out that for an anisotropic star the constraint on the surface gravitational 
red--shift is $z_{s}=5.211$. As can be seen the obtained results are bounded by this value. Moreover, we have checked the impact of parameter $\chi$ on the lower and upper bounds of the surface gravitational red--shift (see table \ref{table33}) and compactness factor (see table \ref{table44}). It is observed that a increasing $\chi$ in magnitude dismiss both $z_{s}$ and $u$ (the lower and upper limits).

\begin{table}[H]
	\centering
	\caption{Lower and upper bounds for the mass-radius ratio and the mass-radius relation for $\chi=0.008$ and values mentioned in table \ref{table1}}\label{table4} 
\begin{tabular}{@{}ccccc@{}}
\hline
Strange & Lower bound & ~\ Mass--Radius ratio & Upper bound   \\
 Star  & $\frac{Q^{2}\left(18R^{2}+Q^{2}\right)}{2R^{2}\left(12R^{2}+Q^{2}\right)}$ & $\frac{M}{R}$ & $\frac{2R^{2}+3Q^{2}+2R\sqrt{R^{2}+3Q^{2}}}{9R^{2}}$  \\ \hline

SMC X--1 ~\ \cite{r47} & 0.0134841&0.1846&0.45636 \\ \hline

LMC X--4 ~\ \cite{r47}& 0.0181804 &0.2152 & 0.46047  \\ \hline
\end{tabular}
\end{table}

\begin{table}[H]
	\centering
	\caption{Lower and upper bounds for the mass-radius ratio and the mass-radius relation for different values of $\chi$ and values mentioned in table \ref{table1}}\label{table44} 
\begin{tabular}{@{}ccccc@{}}
\hline
Strange & $\chi$& Lower bound & &  Upper bound   \\
 Star  && $\frac{Q^{2}\left(18R^{2}+Q^{2}\right)}{2R^{2}\left(12R^{2}+Q^{2}\right)}$  && $\frac{2R^{2}+3Q^{2}+2R\sqrt{R^{2}+3Q^{2}}}{9R^{2}}$  \\ \hline

SMC X--1 ~\ \cite{r47} & 0.2& 0.002798&&0.44693 \\ \hline
SMC X--1 ~\ \cite{r47} & 1.2& 0.000546&&0.44493 \\ \hline
LMC X--4 ~\ \cite{r47}& 0.2 &0.003797 && 0.44781  \\ \hline
LMC X--4 ~\ \cite{r47}& 1.2 &0.000741 && 0.44510  \\ \hline
\end{tabular}
\end{table}

\section{Concluding Remarks}\label{sec7}

In this article we have obtained a well--behaved interior solution describing compact objects such as neutron and quark stars. The main ingredient of this model is a charged anisotropic matter distribution in the stellar interior. To obtain this toy model we have closed the Einstein--Maxwell set of equations by using the class I approach and a link between the anisotropy factor and the electric field. The class I methodology provides the geometry of inner space-time once one of the metric potentials is imposed. In this opportunity we have selected the $g_{rr}$ potential corresponding to the isotropic Buchdhal solution. This choice is well motivated since it is free from physical and mathematical singularities. Placing this metric potential into Eq. (\ref{eq20}) the temporal metric component is determined, completing the geometrical description of the problem. On the other hand, to obtain the full energy--momentum tensor (\ref{eq3}), instead of impose an equation of state, we have established a relation between the anisotropy factor ($\Delta>0$) and the electric field (\ref{eq26}). This relation is mediated by a dimensionless parameter, namely $\chi$. As $\chi$ increases in magnitude the electric field also increases, while the anisotropy factor decreases and vice versa. The structure of this link discards negative values for $\chi$ in order to assure $\Delta>0$ everywhere. The main features characterizing the model \i.e, the density $\rho$, radial pressure $p_{r}$ and transverse pressure $p_{\perp}$ satisfy all the requirement to represent a well behaved stellar interior solution. This is corroborated in Fig. \ref{fig1}, where it is clear that these thermodynamic quantities are positive defined and decreasing functions with increasing radial coordinate $r$. In the same figure, in upper left panel, the metric potentials are exhibited. It is shown that both potentials coincide at the boundary of the object indicating that the junction condition procedure with the exterior Reissner--Nordstr\"om space-time is correct. It is worth mentioning that the energy--momentum tensor satisfies all energy conditions (see Fig. \ref{fig2} for more details), thus the matter distribution is well behaved and admissible from the physical point of view. Respect to the electric properties, they behaved as expected \i.e, null at the center of the structure and positive defined and increasing function within the stellar interior. It should be noted that the order of magnitude of the central density, the electric field and electric charge evaluated at the boundary of the compact star are in complete agreement with the values reported for charged quark stars (see tables \ref{table2} and \ref{table3}). In addition, we have obtained in table \ref{table33} the values for the electric field and electric charge for different values of the parameter $\chi$, and as expected for increasing $\chi$ the electric properties decrease. 

The hydrostatic balance and stability of the system, under the action of the hydrostatic $F_{h}$, gravitational $F_{g}$, anisotropic $F_{a}$ and electric $F_{e}$ gradients was analyzed. The configuration remains in hydrostatic balances under the mentioned gradients. In this regard the electric gradient play an important role to counteract the gravitational one, avoiding a collapse into a point singularity (refer to Fig. \ref{fig4}). As mentioned before, if $\chi$ decreases in magnitude then the anisotropic gradient $F_{a}$ increases, being more relevant than $F_{e}$ in this process. However, regardless of which gradient dominates, they all help to avoid collapse against the gravitational gradient. We studied the stability of the hydrostatic balance from three different schemes, namely: i) relativistic adiabatic index, ii) Harrison--Zeldovich--Novikov and iii) Abreu's et. al criterion. As can be seen in Figs. \ref{fig5} and \ref{fig6} the system is stable in all frames. In table \ref{table2} it is appreciated that the central relativistic adiabatic index overcome the critical value, and the left panel in Fig. \ref{fig6} shows that the matter distribution meets causality condition along the principal directions of the fluid sphere. Finally, we have studied the impact of electric properties on the macro physical observables \i.e, total gravitational mass, mass--radius ratio and surface gravitational red--shift. The trend of this quantities are depicted in Fig. \ref{fig7}. Besides, in tables \ref{table33} and \ref{table44} is shown the impact of parameter $\chi$ on these important observables and also on the electric properties. So, taking into account all these things, we can conclude that the obtained model could represent charged anisotropic compact objects.

\section*{Acknowledgements}
Y. Gomez-Leyton thanks the financial support by the CONICYT PFCHA/DOCTORADO-NACIONAL/$2020$- $21202056$.
F. Tello-Ortiz thanks the financial support by the CONICYT PFCHA/DOCTORADO-NACIONAL/$2019$-$21190856$ and projects ANT-$1756$ and SEM $18-02$ at the Universidad de Antofagasta, Chile. F. Tello-Ortiz thanks
the PhD program Doctorado en Física mención en Física
Matemática de la Universidad de Antofagasta for continuous support and encouragement.

\appendix 

\section{The full energy--momentum tensor}\label{A}

{In this appendix we show how the get the energy--momentum tensor given by Eq. (\ref{eq3}). The derivation provided here is the classical one, that is, we start from the Einstein--Hilbert action minimally coupled to the Maxwell electromagnetic field including both the interaction and particle contributions \cite{weinberg}. So we have
\begin{equation}\label{A1}
S_{\text{Total}}=S_{\text{E--H}}+S_{\text{EM}}+S_{\text{Int}}+S_{\text{Part}},    
\end{equation}
where each action is given by 
\begin{eqnarray}\label{A2}
S_{\text{E--H}}&=&\frac{1}{16\pi}\int d^{4}x \sqrt{-g} R, \\ \label{A3}  
S_{\text{EM}}&=&-\frac{1}{16\pi}\int d^{4}x\sqrt{-g}F_{\mu\nu}F^{\mu\nu}, \\ \label{A4}
S_{\text{Int}}&=&-\sum_{n}e_{n}\int dp \frac{x^{\mu}_{n}}{dp}A_{\mu}, \\ \label{A5}
S_{\text{Part}}&=&-\sum_{n}m_{n}\int dp\left[-g_{\mu\nu}\frac{dx^{\mu}}{dp}\frac{dx^{\nu}}{dp}\right]^{1/2}.
\end{eqnarray}
It worth mentioning that $g_{\mu\nu}$ and $A_{\mu}$ are functions of $x^{\mu}(p)$ being $p$ some quantity parametrizing  the particle trajectories. Furthermore, $m_{n}$ and $e_{n}$ are the mass and electric charge for each point particles. Now, variations with respect to $A_{\mu}$ and $x^{\mu}$ lead to the following field equations for he electromagnetic field and the point particles
\begin{eqnarray}\label{A6}
\nabla_{[\alpha}F_{\beta\gamma]}&=&0, \\ \label{A7}
\partial_{\mu}\left[\sqrt{-g}F^{\mu\nu}\right]&=&4\pi\sum_{n}e_{n}\int \delta^{4}\left(x-x_{n}\right)\frac{dx^{\nu}_{n}}{d\tau_{n}}d\tau_{n},
\end{eqnarray}
where it is clear from (\ref{A7}) that
\begin{equation}\label{A8}
\sqrt{-g}J^{\nu}\equiv \sum_{n}e_{n}\int \delta^{4}\left(x-x_{n}\right)\frac{dx^{\nu}_{n}}{d\tau_{n}}d\tau_{n}.    
\end{equation}
Moreover, to satisfy (\ref{A6})--(\ref{A7}) the tensor $F_{\mu\nu}$ is defined as $F_{\mu\nu}=\partial_{\mu}A_{\nu}-\partial_{\nu}A_{\mu}$.
Next, for the point particles one has
\begin{equation}\label{A9}
\frac{d^{2}x^{\mu}_{n}}{d\tau^{2}_{n}}+\Gamma^{\mu}_{\nu\lambda}\frac{dx^{\lambda}_{n}}{d\tau_{n}}\frac{dx^{\nu}_{n}}{d\tau_{n}}=\left(\frac{e_{n}}{m_{n}}\right)F^{\mu}_{\ \nu}\frac{dx^{\nu}_{n}}{d\tau_{n}},   
\end{equation}
and $d\tau_{n}$ is given by
\begin{equation}\label{A10}
d\tau_{n}\equiv \left[-g_{\mu\nu}dx^{\mu}dx^{\nu}\right]^{1/2}.   
\end{equation}
Variation with respect to $g_{\mu\nu}$ leads to the Einstein tensor
\begin{equation}\label{A11}
G_{\mu\nu}\equiv R_{\mu\nu}-\frac{R}{2}g_{\mu\nu},    
\end{equation}
whilst the remaining terms conform the energy--momentum tensor 
\begin{equation}\label{A12}
T^{\mu\nu}=\frac{1}{\sqrt{-g}}\sum_{n}m_{n}\int d\tau_{n}\frac{dx^{\mu}_{n}}{d\tau_{n}}\frac{dx^{\nu}_{n}}{d\tau_{n}} \delta^{4}\left(x-x_{n}\right)+\frac{1}{4\pi}\left(\frac{1}{4}g^{\mu\nu}F_{\alpha\beta}F^{\alpha\beta}-F^{\mu\alpha}F^{\nu}_{\ \alpha}\right).     
\end{equation}
The first term in (\ref{A12}), under certain assumptions could represent any type of matter distribution \cite{sh}. In this case this term is describing an imperfect or anisotropic fluid, expressed by
\begin{equation}\label{A13}
\frac{1}{\sqrt{-g}}\sum_{n}m_{n}\int d\tau_{n}\frac{dx^{\mu}_{n}}{d\tau_{n}}\frac{dx^{\nu}_{n}}{d\tau_{n}} \delta^{4}\left(x-x_{n}\right)= \left({\rho} +{\bar{p}}_{\perp}\right)\,\chi^{\mu}\,\chi^{\nu}-g^{\mu\nu}\,{\bar{p}}_{\perp}+\left({\bar{p}}_{r}-{\bar{p}}_{\perp}\right)\,u^{\mu}\,u^{\nu}. 
\end{equation}
Then, the field equations for the gravitational sector are
\begin{equation}\label{A14}
G_{\mu\nu}=8\pi T_{\mu\nu}. 
\end{equation}
By virtue of Bianchi's identities one has
\begin{equation}\label{A15}
\nabla_{\mu}G^{\mu\nu}=0 \Rightarrow \nabla_{\mu}T^{\mu\nu}=0. 
\end{equation}
So, from Eqs. (\ref{A12})--(\ref{A13}) it is not hard to obtain 
\begin{equation}\label{A16}
\nabla_{\mu}T^{\mu\nu}=\nabla_{\mu}\left[\left({\rho} +{\bar{p}}_{\perp}\right)\,\chi^{\mu}\,\chi^{\nu}-g^{\mu\nu}\,{\bar{p}}_{\perp}+\left({\bar{p}}_{r}-{\bar{p}}_{\perp}\right)\,u^{\mu}\,u^{\nu}\right]-F^{\nu}_{\ \alpha}J^{\alpha}.    
\end{equation}
In obtaining the second member in the right hand side of (\ref{A16}) we have employed the Eqs. (\ref{A6})--(\ref{A7}). Now, combining Eqs. (\ref{A8}), (\ref{A9}) and (\ref{A13}) one arrives to
\begin{equation}\label{A17}
 \nabla_{\mu}\left[\left({\rho} +{\bar{p}}_{\perp}\right)\,\chi^{\mu}\,\chi^{\nu}-g^{\mu\nu}\,{\bar{p}}_{\perp}+\left({\bar{p}}_{r}-{\bar{p}}_{\perp}\right)\,u^{\mu}\,u^{\nu}\right]=F^{\nu}_{\ \alpha}J^{\alpha}.   
\end{equation}
Thus, inserting (\ref{A17}) into (\ref{A16}) one gets $\nabla_{\mu}T^{\mu\nu}=0$ as desired. As can be seen, the whole energy--momentum tensor is conserved \i.e, electromagnetic field (containing the interaction) plus point particles contribution.}

\end{document}